\documentclass[5p,times,twocolumn]{elsarticle}

\usepackage[T1]{fontenc}
\usepackage[utf8]{inputenc}
\usepackage{microtype}
\usepackage{amsmath}
\usepackage{graphicx}
\usepackage{booktabs}
\usepackage{array}
\usepackage{multirow}
\usepackage[table]{xcolor}
\usepackage[hidelinks]{hyperref}
\usepackage{xurl}
\usepackage{tabularx}
\usepackage{makecell}

\definecolor{dimesblue}{RGB}{153,213,225}
\definecolor{exposureorange}{RGB}{228,126,35}
\definecolor{samplesgreen}{RGB}{112,173,71}
\definecolor{angledpurple}{RGB}{211,193,235}
\definecolor{tiltedorange}{RGB}{255,220,158}

\newcolumntype{C}[1]{>{\centering\arraybackslash}m{#1}}
\newcolumntype{Y}{>{\raggedright\arraybackslash}X}

\newcommand{\ang}[1]{%
  \begingroup\setlength{\fboxsep}{1pt}\colorbox{angledpurple}{\strut #1}\endgroup%
}
\newcommand{\tilt}[1]{%
  \begingroup\setlength{\fboxsep}{1pt}\colorbox{tiltedorange}{\strut #1}\endgroup%
}

\journal{Nuclear Materials and Energy}

\begin{document}
\begin{frontmatter}
\title{Comparative qualification of advanced plasma-facing materials for fusion pilot plants through public- and private-sector experiments in DIII-D}

\author[1]{F. Effenberg\corref{cor1}}
\cortext[cor1]{Corresponding author}
\ead{feffenbe@pppl.gov}
\author[2]{J.D. Coburn}
\author[3]{L. Cappelli}
\author[4]{S.D. Corah}
\author[5]{A. Grandhi}
\author[6]{J. Guterl}
\author[7]{C. Hirst}
\author[8]{M. Jackson}
\author[7]{D. Kohler}
\author[2]{R. Kolasinski}
\author[9]{E. Martinez-Loran}
\author[10]{R. Meston}
\author[11]{L. Nuckols}
\author[12]{A. Ottaviano}
\author[13]{\v{Z}. Popovi\'c}
\author[5]{S. Tsurkan}
\author[10]{T. Van Volkenburg}
\author[5]{D. Velazquez}
\author[14]{A. Zuniga}
\author[4]{A. Bhattacharya}
\author[9]{J. Boedo}
\author[16]{K. Christian}
\author[13]{G. Dose}
\author[9]{E. Hollmann}
\author[7,15]{M. Ialovega}
\author[14]{F.M. Laggner}
\author[8]{S. Levine}
\author[7]{M.X. Navarro-Gonzalez}
\author[8]{A. Shone}
\author[16]{L.L. Snead}
\author[16]{D. Sprouster}
\author[16]{J. Trelewicz}
\author[13]{J. Barr}
\author[2]{R. Hood}
\author[17]{C. Lasnier}
\author[11]{Y. Lin}
\author[18]{V. Mazon}
\author[19]{J. Ren}
\author[11]{G. Ronchi}
\author[9]{D. Rudakov}
\author[17]{D. Truong}
\author[2]{C. Tsui}
\author[13]{S. Zamperini}
\author[11]{W. Zhong}
\author[13]{T. Abrams}
\author[13]{A.W. Leonard}
\author[13]{R. Maurizio}
\address[1]{Princeton Plasma Physics Laboratory, 100 Stellarator Road, Princeton, NJ 08540, USA}
\address[2]{Sandia National Laboratories, 7011 East Avenue, Livermore, CA 94550, USA}
\address[3]{Oak Ridge Associated Universities, 100 ORAU Way, Oak Ridge, TN 37830, USA}
\address[4]{School of Metallurgy and Materials, University of Birmingham, Pritchatts Road, Edgbaston, Birmingham B15 2TT, United Kingdom}
\address[5]{Avalanche Energy Designs, Inc., 9100 E Marginal Way S, Tukwila, WA 98108, USA}
\address[6]{OpenStar Technologies, Wellington, New Zealand}
\address[7]{University of Wisconsin--Madison, Madison, WI 53706, USA}
\address[8]{Tokamak Energy Ltd, 173 Brook Drive, Milton Park, Abingdon, Oxfordshire OX14 4SD, United Kingdom}
\address[9]{Center for Energy Research, University of California San Diego, 9500 Gilman Drive, MC 0417, La Jolla, CA 92093-0417, USA}
\address[10]{Helion Energy, Inc., 1415 75th Street SW, Everett, WA 98203, USA}
\address[11]{Oak Ridge National Laboratory, 1 Bethel Valley Road, Oak Ridge, TN 37830, USA}
\address[12]{Thea Energy, Inc., 1 Eastern Road, Suite 3-04, Kearny, NJ 07032, USA}
\address[13]{General Atomics, 3550 General Atomics Court, San Diego, CA 92121-1122, USA}
\address[14]{Department of Nuclear Engineering, North Carolina State University, 915 Partners Way, Room 4131, Raleigh, NC 27607, USA}
\address[15]{GenF, 2 Avenue Gay Lussac, 78990 Elancourt, France}
\address[16]{Stony Brook University, 100 Nicolls Road, Stony Brook, NY 11794, USA}
\address[17]{Lawrence Livermore National Laboratory, 7000 East Avenue, Livermore, CA 94550-9698, USA}
\address[18]{IMT Mines Albi, Campus Jarlard, 81013 Albi CT Cedex 09, France}
\address[19]{University of Tennessee, Knoxville, Knoxville, TN 37996, USA}

\begin{abstract}
A coordinated DIII-D campaign involving four private fusion developers and 12 participating institutions comparatively assessed 44 advanced plasma-facing materials, including the first DIII-D divertor exposure of neutron-irradiated W, to support fusion pilot plant wall and divertor material qualification. Samples were exposed using the Divertor Materials Evaluation System (DiMES) under Ohmic, L-mode, and H-mode conditions with edge-localized modes, at 0.2--2.5~MW~m\(^{-2}\) on flush geometries and 10--15~MW~m\(^{-2}\) on 10\(^{\circ}\) angled geometries. Engineered tungsten architectures retained integrity; long-fiber Wf/W showed the clearest crack-arrest behavior. W--Re and K-doped W showed near-ITER-W-like responses, while additively manufactured W--Ta showed heat-flux-sensitive mass losses of 0.64~mg for the flat sample and 2.19--2.87~mg for angled samples. After irradiation to 0.3~dpa at 550~\(^{\circ}\)C, neutron-irradiated ITER-grade W retained 2.8 times more deuterium than pristine W, while TiB\(_2\) showed the lowest D\(_2\) release in the Ohmic set. VTaHfMo was the most stable refractory multi-principal-element alloy. NbC and \((\mathrm{Nb}_{0.5}\mathrm{Ta}_{0.5})\mathrm{C}\) retained integrity with 0.02--0.03~mg mass loss, whereas ZrC lost 7~mg. CVD SiC retained macroscopic integrity but exhibited an effective Si erosion yield of 0.5, about 5--10 times above prior DIII-D trends. Renewable boron pebble rods underwent plasma-driven recession; 13\% of released boron was ionized near the outer strike point and up to 50\% was recovered locally. Initial in-situ Cr gross-erosion measurements yielded effective erosion yields of order \(10^{-2}\). Together, these results provide cross-material benchmarks for fusion pilot plant down-selection and future AI/ML-assisted plasma-facing-material development.
\end{abstract}

\begin{keyword}
fusion reactor materials \sep
plasma-facing component qualification \sep
tungsten-based materials \sep
additive manufacturing \sep
neutron irradiation \sep
renewable plasma-facing materials \sep
fusion materials database
\end{keyword}
\end{frontmatter}

\section{Introduction}

The development of plasma-facing materials (PFMs) for fusion pilot plants (FPPs) requires validation under conditions where high heat and particle fluxes, transient loading, erosion, fuel retention, and neutron-induced material modification act simultaneously. The coupled nature of these processes governs material lifetime and plasma compatibility and must be assessed in integrated environments representative of reactor operation \cite{Philipps2011,Linke2019,Linsmeier2017}.

High-Z materials, in particular tungsten and its alloys, are widely considered the leading candidates for divertor and first-wall applications due to their low sputtering yields, high melting temperatures, and comparatively favorable tritium retention characteristics \cite{Roth2009,Coenen2020}. Control of long-term fuel retention has been a central driver in the transition away from carbon-based plasma-facing components, where co-deposition leads to significant tritium accumulation \cite{Roth2009,Brezinsek2022}. In ITER, this motivated the decision to begin operation with a full-tungsten divertor, and the 2024 re-baseline also replaced the beryllium first wall with tungsten~\cite{Pitts2019,Pitts2025}. Experimental validation of this transition was provided by the JET ITER-like wall, where carbon impurity levels and long-term fuel retention were reduced by more than an order of magnitude, albeit with increased constraints on operational space due to tungsten influx control \cite{Brezinsek2015}. Consistent with these findings, tungsten has been adopted as the baseline material in ITER, SPARC, and European DEMO concepts \cite{Pitts2019,You2021,Creely2023,Pitts2025,Richiusa2022}, and is now being used or planned for in devices such as ASDEX Upgrade, WEST, KSTAR, DIII-D, EAST and W7-X \cite{Neu2026,Bucalossi2014,Ko2024,Abrams2024,Wan2019,Fellinger2023}.

Despite these advantages, tungsten performance is limited by intrinsic brittleness and sensitivity to cyclic and transient heat loads, leading to crack formation, recrystallization, and surface degradation under reactor-relevant conditions, particularly when combined with neutron irradiation. These limitations have driven the development of advanced tungsten-based materials, including alloyed, dispersion-strengthened, fiber-reinforced, and additively manufactured microstructures with controlled grain characteristics.

In parallel, alternative plasma-facing material concepts, including refractory multi-principal-element alloys (MPEAs), advanced ceramic and ultra-high-temperature ceramic coatings, silicon carbide-based composites, chromium-based coatings, and renewable or continuously replenished surfaces based on lithium, tin, and boron-rich materials, are being explored to extend performance under extreme plasma heat and particle loads and to mitigate limitations associated with cracking, transient damage, erosion, and fuel retention \cite{Kessel2019,Morbey2025,Nygren2016}. These approaches broaden the PFM design space for future fusion pilot plants, encompassing both large D-T tokamaks and more compact concepts, including compact/high-power-density tokamaks \cite{Menard2022}, low-aspect-ratio tokamak power-plant studies \cite{Willis2026}, and stellarator power-plant concepts \cite{Swanson2025}, as well as pulsed fusion concepts with distinct first-wall loading requirements. In this context, the PFMs examined in the present study are evaluated under DIII-D divertor conditions as candidates to manage high heat and particle fluxes, control fuel retention, and extend component lifetime across this spectrum of FPP designs. Recent high-throughput PFM screening further motivates experimental benchmarking across a broad candidate-material space \cite{Fedrigucci2024}.

A central limitation of many plasma-facing material studies is the separation of plasma exposure, neutron irradiation, and materials characterization, despite reactor-relevant performance being governed by their coupled effects on erosion, surface evolution, thermomechanical degradation, and defect accumulation. Consistent with recent fusion materials roadmaps, advancing candidate plasma-facing materials requires integrated testing approaches that assess material behavior under simultaneous thermal, particle, and neutron irradiation loading while generating the data needed for predictive model development and technology maturation \cite{Linke2019,FMCC2025, DOEFSTRoadmap2026}.

The DIII-D National Fusion Facility, using the Divertor Materials Evaluation System (DiMES) \cite{Wong1992,Rudakov2017}, provides a unique capability for comparative assessment of candidate plasma-facing materials under controlled tokamak conditions. Building on the reference exposure framework and initial screening of 17 candidate materials during the first ``FPP Materials Thrust'' campaign \cite{Coburn2026}, the second campaign expanded the material portfolio and institutional participation, including four private fusion developers for the first time within the Thrust. The private-sector sample sets reflected concept-specific PFM requirements: Helion Energy's pulsed fusion approach motivated low-Z ceramic first-wall candidates for transient, repetitively cycled heat loading; Thea Energy's stellarator divertor studies motivated renewable boron surfaces for localized exhaust handling and material recovery; Avalanche Energy's compact fusion concept motivated refractory MPEAs for localized high-flux components; and Tokamak Energy's low-aspect-ratio tokamak program motivated collaborative evaluation of advanced refractory-metal PFCs for compact divertor conditions \cite{Willis2026}. A total of 44 additional candidate materials were evaluated, extending the initial screening to neutron-irradiated W, new engineered W architectures, W alloys and doped W, refractory alloys and MPEAs, advanced ceramics, and renewable B surfaces. Beyond material survivability, the present study quantifies D$_2$ release from neutron-irradiated W, effective Cr and Si erosion yields, and B-pebble release and recovery, while identifying crack-arrest mechanisms and damage localization in engineered W and composition-dependent degradation of refractory alloys and ceramics. Materials were characterized before and after exposure under established DiMES reference scenarios, supplemented by in-situ diagnostics where available. Quantitative comparisons are restricted to sufficiently matched exposure conditions; comparisons across different plasma scenarios or sample geometries are interpreted qualitatively.
\begin{table}[htbp]
\centering
\caption{Representative reference plasma conditions used for DiMES material exposures. Values are representative at or near the DiMES radial location; \(q_{\perp}\) is normal to the exposed sample surface, while the listed \(\Gamma_{\perp}\) is normal to the nominal divertor target for flush and Ohmic reference conditions, and to the inclined sample for angled H-mode. The particle-flux values represent peak values from reference discharges. $T_{\mathrm{exposure}}$ denotes the DiMES dwell time per discharge. The Ohmic scenario was used for HFIR-compatible 5\(^{\circ}\) angled samples, L-mode for selected material exposures, and rastered H-mode for most comparative material exposures (see Table~\ref{tab:02}).}
\label{tab:01}
\small
\resizebox{\linewidth}{!}{%
\begin{tabular}{lllllll}
\toprule
Scenario & Sample Type & $q_{\perp}$ [$\mathrm{MW}\,\mathrm{m}^{-2}$] & $n_e$ [$\mathrm{m}^{-3}$] & $T_e$ [$\mathrm{eV}$]& $\Gamma_{\perp}$ [$\mathrm{m}^{-2}\,\mathrm{s}^{-1}$] & $T_{\mathrm{exposure}}$ [s] \\
\midrule
Ohmic & $5^{\circ}$ angled & 0.8--1.3 & $2\times10^{19}$ & 17 & $1.7\times10^{22}$ & 1--2.5 \\
L-mode & Flush & 1.3--1.7 & $1.5\times10^{19}$ & 25--35 & $1.6\times10^{22}$ & 2--3 \\
\multirow{2}{*}{\shortstack[l]{H-mode,\\rastering}} & Flush & 1.6--2 & \multirow{2}{*}{$1.2\times10^{19}$} & \multirow{2}{*}{20--30} & $1.5\times10^{22}$ & 2 \\
& $10^{\circ}$ angled & 12--15 & & & $1.1\times10^{23}$ & 2 \\
\bottomrule
\end{tabular}%
}
\end{table}

These coordinated material exposure experiments support the maturation of low-technological readiness level (TRL) plasma-facing materials by exposing them to reactor-relevant plasma conditions that simultaneously impose high heat and particle fluxes, reactor-relevant impact energies and angles, strong magnetic fields, elevated surface temperatures, cyclic thermal loading, and erosion-redeposition processes \cite{FMCC2025}. For the first time, neutron-irradiated tungsten specimens were also incorporated into DIII-D material exposure studies. Although exposure durations were typically a few seconds and active cooling is absent, the resulting environment cannot be readily reproduced in linear plasma devices and provides an important intermediate step between laboratory-scale testing and component-level development. Following the framework proposed in \cite{Coburn2026}, these experiments contribute to the advancement of candidate plasma-facing materials from laboratory proof-of-concept toward TRL 3--4 fusion-relevant materials evaluation.

The material exposure experiments evaluated key plasma-material interactions (PMI) performance metrics, including erosion and material loss, fuel retention, thermomechanical response, surface integrity, and plasma compatibility. The resulting dataset supports comparisons within matched exposure sets and identifies degradation mechanisms relevant to DEMO and future FPPs, including cracking, melting, blistering, recrystallization, selective erosion, irradiation-enhanced retention, and controlled material release under transient plasma loading. Because the samples were exposed under common DiMES reference scenarios and evaluated using comparable PMI metrics, the resulting dataset supports predictive materials modeling, future MatDB4Fusion archiving, and data-driven PFM screening for DEMO and fusion pilot plants.

\section{Experimental setup and plasma scenarios for comparative materials qualification}
\begin{table}[!t]
\centering
\tiny
\setlength{\tabcolsep}{2pt}
\renewcommand{\arraystretch}{1.08}

\caption{List of DiMES holders, samples, and exposure conditions. Exposure column values correspond to the scenarios listed in Table~\ref{tab:01}:
Ohmic = stationary ohmic, L = stationary L-mode, H = stationary H-mode, H (raster) = rastering H-mode, and H+VDE = stationary H-mode discharge ending with an intentional downward vertical displacement event (VDE). Purple entries indicate \(10^{\circ}\) angled samples; light-orange entries indicate samples tilted \(5^{\circ}\) towards the incident plasma. Unshaded entries in seven-button holders are flush-mounted. Samples are listed in holder-position order. Neutron-irradiated samples are indicated in bold with an *.}
\label{tab:02}

\begin{tabularx}{\columnwidth}{|C{0.55cm}|C{1.05cm}|Y|}
\hline
\cellcolor{dimesblue}\rule[-0.6ex]{0pt}{3.2ex}\textbf{\#} &
\cellcolor{exposureorange}\rule[-0.6ex]{0pt}{3.2ex}\makecell{\textbf{Expo-}\\[-0.2ex]\textbf{sure}} &
\cellcolor{samplesgreen}\rule[-0.6ex]{0pt}{3.2ex}\centering\arraybackslash\textbf{Samples} \\
\hline

\cellcolor{dimesblue}1 &
\cellcolor{exposureorange}L &
W; W; Ta CS; W-Ta CS; LPBF-W; LPBF-W; LPBF-W \\
\hline

\cellcolor{dimesblue}2 &
\cellcolor{exposureorange}H (raster) &
W; W; Ta CS; W-Ta CS; LPBF-W; Ta; W-Ta CS \\
\hline

\cellcolor{dimesblue}3 &
\cellcolor{exposureorange}H (raster) &
\ang{W-K}; \ang{W-K}; \ang{W-Re}; \ang{W}; W-K; W-Re; W-K \\
\hline

\cellcolor{dimesblue}4 &
\cellcolor{exposureorange}H (raster) &
\ang{W-Ti-Cr}; \ang{EB-PBF-W-Re}; \ang{EB-PBF-W}; NbTaC; W-Ti-Cr; NbC; ZrC \\
\hline

\cellcolor{dimesblue}5 &
\cellcolor{exposureorange}H (raster) &
\ang{B\(_4\)C}; B\(_4\)C; \ang{Si\(_3\)N\(_4\)}; Si\(_3\)N\(_4\); \ang{SiC CVD}; \textbf{SiC CVD*}; SiC CVD \\
\hline

\cellcolor{dimesblue}6 &
\cellcolor{exposureorange}H (raster) &
SiC; SiC CVD; SiC CVI; SiC, O2; SiC; SiC CVD; SiC CVI \\
\hline

\cellcolor{dimesblue}7 &
\cellcolor{exposureorange}H (raster) &
\ang{LPBF W--10Ta}; \ang{LPBF W--10Ta}; W-Ta; \ang{W-K}; W-K; W; W \\
\hline

\cellcolor{dimesblue}8 &
\cellcolor{exposureorange}H (raster) &
\ang{LPBF W--10Ta}; \ang{LPBF W--10Ta}; W-Ta; \ang{W-K}; W-K; W-K; W \\
\hline

\cellcolor{dimesblue}9 &
\cellcolor{exposureorange}H (raster) &
VTaHfMo; HfZrTiTa; ZrTiCr; ZrTiV; ZrTiHf; ZrTiW; ZrTiTa \\
\hline

\cellcolor{dimesblue}10 &
\cellcolor{exposureorange}L, H &
B pebble rod, 2 mm height \\
\hline

\cellcolor{dimesblue}11 &
\cellcolor{exposureorange}L &
B pebble rod, 5 mm height \\
\hline

\cellcolor{dimesblue}12 &
\cellcolor{exposureorange}H + VDE &
B pebble rod, 2 mm height \\
\hline

\cellcolor{dimesblue}13 &
\cellcolor{exposureorange}Ohmic &
\tilt{W}; \tilt{W-K}; \tilt{TiB\(_2\)}; \tilt{\textbf{W*}} \\
\hline

\cellcolor{dimesblue}14 &
\cellcolor{exposureorange}H (raster) &
\ang{EB-PBF-W}; \ang{EB-PBF-W}; SiC; WC; (W\(_{0.5}\)Si\(_{0.5}\))C; W; W-Ti-Cr \\
\hline

\cellcolor{dimesblue}15 &
\cellcolor{exposureorange}H (raster) &
\ang{EB-PBF-W}; \ang{EB-PBF-W}; EB-PBF-W; EB-PBF-W;
EB-PBF-W; EB-PBF-W; \ang{W} \\
\hline

\cellcolor{dimesblue}16 &
\cellcolor{exposureorange}H (raster) &
\ang{EB-PBF-W}; \ang{EB-PBF-W}; EB-PBF-W; EB-PBF-W;
EB-PBF-W; EB-PBF-W; \ang{W} \\
\hline

\cellcolor{dimesblue}17 &
\cellcolor{exposureorange}\makecell{H,\\raster} &
\ang{$\mu$-W}; \ang{$\mu$-W}; \ang{Wf/W long}; Wf/W short; Cr; Wf/SiCf/W; Cr \\
\hline

\cellcolor{dimesblue}18 &
\cellcolor{exposureorange}H (raster) &
W; Ta CS; W-Ta CS; Ta-W CS; LPBF-W; SiC CVD; SiC CVD \\
\hline

\end{tabularx}
\end{table}

All material exposure experiments used the Divertor Materials Evaluation System (DiMES) in the DIII-D National Fusion Facility \cite{Luxon2002,Fenstermacher2022}. The DiMES manipulator allows for plasma exposure of material samples of varying sizes and geometries (up to 50 mm diameter) to plasmas in the lower divertor region. An expansive suite of diagnostics are available for dedicated PMI studies in the lower divertor. Diagnostics with direct views/measurements of DiMES include several spectroscopic diagnostics (Multichordal Divertor Spectrometer (MDS), high-resolution UV and WISE spectrometers), filtered visible imaging (DiMES TV), and IR imaging systems. Additionally, Langmuir probes (LPs), additional IR imaging systems, and Divertor Thomson Scattering (DTS) provide measurements of plasma parameters (electron density ne and temperature Te) and incident heat flux at the DiMES radial location but at varying toroidal locations. References for the DIII-D poloidal cross-section with diagnostic overlays can be found in \cite{Rudakov2017}. PMI experiments typically expose DiMES samples to outer strike point (OSP) plasmas of a lower single null (LSN) magnetic configuration, which can either be swept over DiMES at a specified rate or left stationary over the duration of the discharge. When longer exposure times are needed, reproducible repeat discharges are used, often with a shot cycle of 10-15 minutes depending on magnetic coil cooldown time.
\begin{figure}[!t]
\centering
\includegraphics[alt={Radial heat flux profiles and time-dependent plasma signals for reference DIII-D materials exposure discharges.},width=0.7\linewidth]{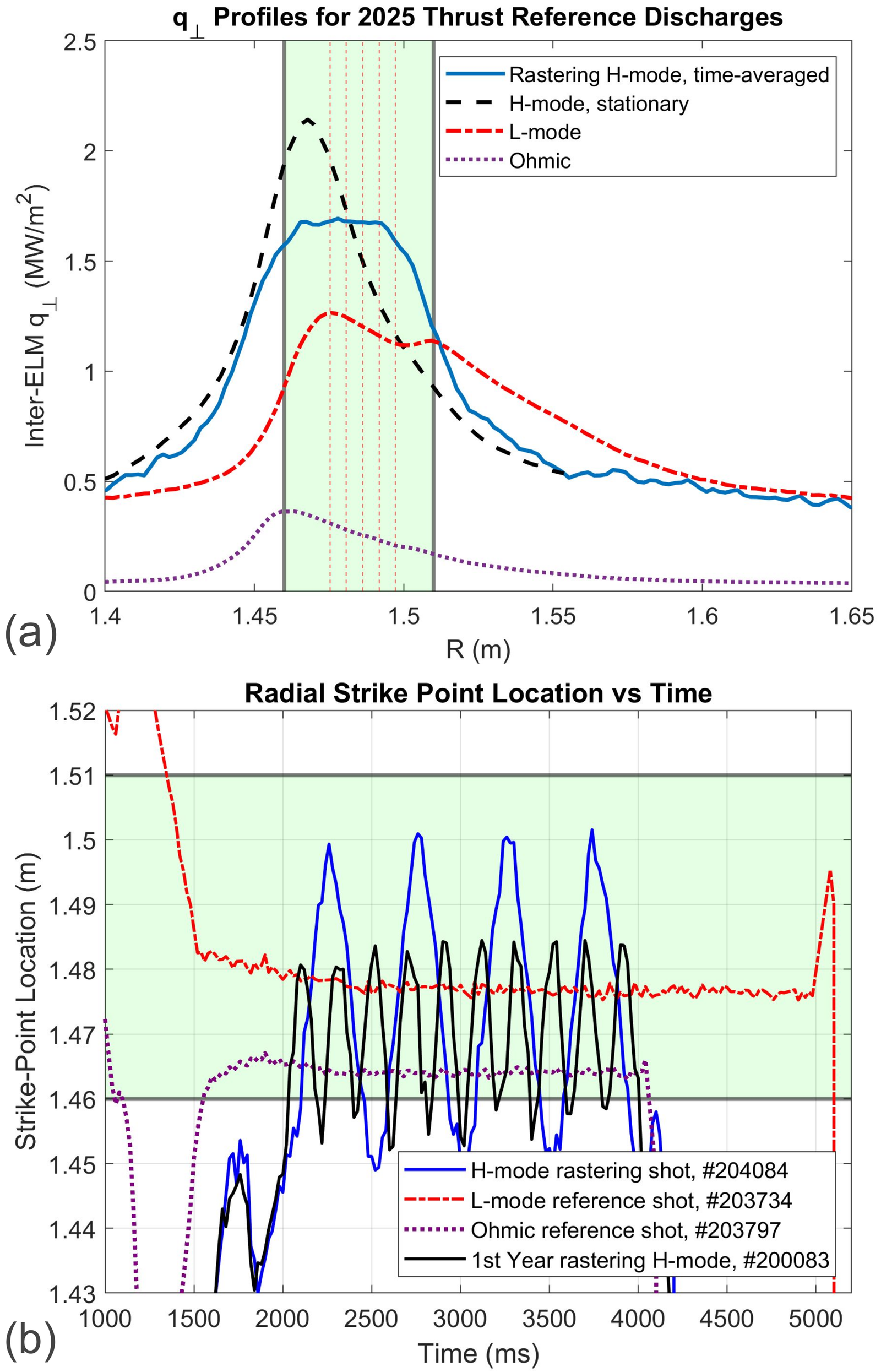}
\caption{(a) Radial heat flux profiles of reference plasma discharges used during DiMES sample exposures, measured by infrared thermography. The DiMES location is indicated in green, with 7-button sample locations indicated by vertical dashed red lines. (b)  Outer strike-point position versus time for the corresponding reference scenarios. The revised H-mode rastering scenario uses a frequency of 2 Hz and a radial sweep of \(\sim\)5 cm. The peak heat-flux location is typically 1–2 cm outboard of the strike-point position.}
\label{fig:01}
\end{figure}

\begin{figure*}[!t]
\centering
\includegraphics[alt={Post-exposure SEM views of engineered tungsten composite architectures including fiber-reinforced tungsten.},width=0.8\linewidth]{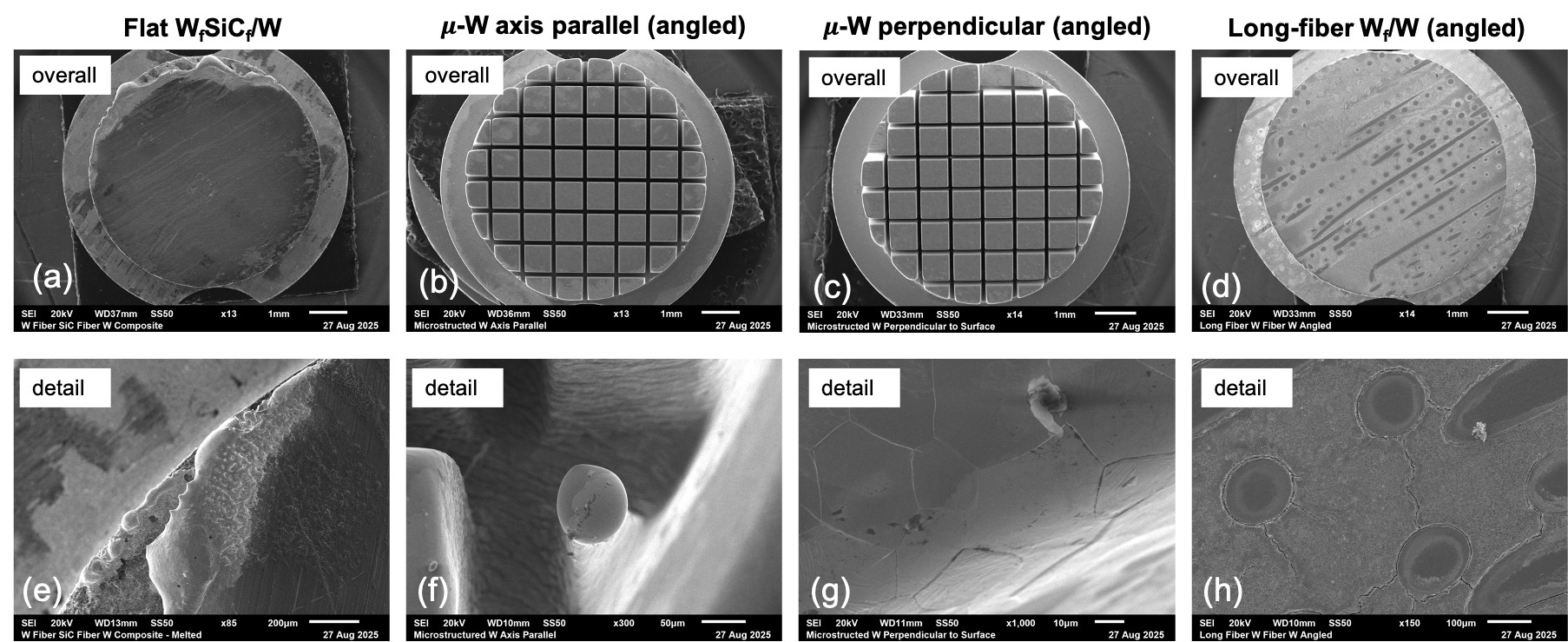}
\caption{Post-exposure SEM images of engineered tungsten architectures exposed in DiMES under ELMy H-mode conditions. Overview images show (a) flush W$_{\mathrm{f}}$/SiC$_{\mathrm{f}}$/W composite, (b) $10^{\circ}$ angled micro-castellated W with castellations parallel to the surface axis, (c) $10^{\circ}$ angled micro-castellated W with castellations perpendicular to the surface axis, and (d) $10^{\circ}$ angled long-fiber W$_{\mathrm{f}}$/W composite. Corresponding higher-magnification images show (e) localized leading-edge melting and resolidification in W$_{\mathrm{f}}$/SiC$_{\mathrm{f}}$/W, (f) localized melting of an individual micro-castellated W column, (g) resolidified grains with intergranular microcracks in micro-castellated W, and (h) surface cracking in long-fiber W$_{\mathrm{f}}$/W, illustrating crack arrest at fiber--matrix interfaces.}
\label{fig:02}
\end{figure*}

Three DiMES holder designs were used to expose samples under the reference Ohmic, L-mode, and rastered H-mode scenarios summarized in Table~\ref{tab:01}. Most experiments used the standard seven-button holder, which accommodated up to seven 6 mm diameter samples in either flush or 10\(^{\circ}\) angled geometry and often included thermocouples for temperature monitoring and modeling support. The 10\(^{\circ}\) geometry increased the effective intercepted heat and particle fluxes by about a factor of seven, from \(q_{\perp}\) \(\approx\) 1.6--2 MW m\(^{-2}\) and \(\Gamma_{\perp}\) \(\approx\) 1.5 \(\times\) 10\(^{22}\) m\(^{-2}\) s\(^{-1}\) for flush H-mode exposure to \(q_{\perp}\) \(\approx\) 12--15 MW m\(^{-2}\) and \(\Gamma_{\perp}\) \(\approx\) 1.1 \(\times\) 10\(^{23}\) m\(^{-2}\) s\(^{-1}\) for angled H-mode exposure, with representative target-parallel fluxes \(\Gamma_{\parallel}\) \(\approx\) 6.2--6.8 \(\times\) 10\(^{23}\) m\(^{-2}\) s\(^{-1}\). Here, \(q_{\perp}\) denotes the sample-normal heat flux, \(\Gamma_{\perp}\) the nominal target-normal particle flux for flush and Ohmic reference conditions and the sample-normal flux for angled H-mode exposures, \(\Gamma_{\parallel}\) the field-aligned particle flux, and \(n_e\) and \(T_e\) the local electron density and temperature. In addition to this standard holder, two custom DiMES heads were designed to accommodate atypical sample geometries and exposure requirements. The first was designed around the thin (\(\sim\)0.5 mm) 6 mm diameter samples used for neutron irradiation studies in the High Flux Isotope Reactor (HFIR) \cite{Cetiner2022}. HFIR enables controlled-temperature tungsten irradiations to DEMO/FPP-relevant damage levels, including \(\sim\)0.5 dpa campaigns on week-scale irradiation times using dedicated capsule designs. These samples do not have a tophat-style edge to help hold samples in place within DiMES, and once neutron irradiated, cannot be further machined. Therefore, a new DiMES used a Titanium-Zirconium-Molybdenum (TZM) plate with thin 0.2 mm lip edges for the samples to press against when installed, leaving them slightly recessed. The entire plasma-facing surface of this holder is inclined 5\(^{\circ}\) towards the incident plasma, which was predicted by modeling to balance the effects of shadowing of samples (by the surrounding 0.2 mm lip) with the predicted increased \(q_{\perp}\) and temperature. A second novel custom holder exposed 1 cm diameter boron pebble rods composed of \(\sim\)1 mm B pebbles in a carbon binder. The rod was surrounded by a boron nitride (BN) insulator for current measurements and a 3~mm deep graphite collection well to recover released B pebbles and fragments. Protrusion height was measured above the well wall, which was aligned with the adjacent divertor tiles (see Section~\ref{sec:boron} and Figure~\ref{fig:16}(a)).

The reference discharge scenarios developed in \cite{Coburn2026} were adapted for the present experiments to allow for compatibility and comparability across material exposure experiments for both campaigns. Reference scenarios for this campaign are summarized in Table~\ref{tab:01} below, with example radial \(q_{\perp}\) profiles and strike-point programs highlighted in Figure~\ref{fig:01}. The Year 2 L-mode exposures used lower neutral beam injection (NBI) power than in Year 1 to reduce heat and particle fluxes and remain below the L–H transition threshold. For the rastering H-mode scenario, see \cite{Coburn2026} for discussion on the development of this scenario. The first year case utilized a raster distance of 3 cm at 5 Hz in order to broaden the \(q_{\perp,avg}\) profile across DiMES. Although this was a great improvement compared to a stationary H-mode case, the scenario did not broaden/flatten \(q_{\perp,avg}\) as much as initially desired. So, the Year 2 scenarios opted for a lower raster frequency of 2 Hz, allowing for better OSP control to achieve an effective 5 cm raster distance. The revised rastering scenario produces a broader, more uniform time-averaged heat-flux profile across the DiMES samples, as shown in Figure~\ref{fig:01}(a). The NBI heating was increased slightly to bring the \(q_{\perp,avg}\) flattop values in-line with those achieved in 2025. See Table~\ref{tab:01} and Figure~\ref{fig:01} of \cite{Coburn2026} for comparison between the L-mode and H-mode reference scenarios. Finally, a low-power plasma discharge using only Ohmic plasma heating was developed explicitly for exposure of the 5\(^{\circ}\) angled DiMES holder for the HFIR samples in order to build up D fluence but remain below the sample irradiation temperature of 550 \(^{\circ}\mathrm{C}\).

Table~\ref{tab:02} lists all of the DiMES holders and samples exposed in the second material testing campaign, along with their exposure conditions referenced to Table~\ref{tab:01}. The EB-PBF W--Re specimen in DiMES \#4 and EB-PBF W specimens in DiMES \#4 and \#14 were supplied by ORNL; the EB-PBF W specimens in DiMES \#15--16 were produced at NCSU. An additional UW--Madison LPBF W specimen was exposed in DiMES \#18. DiMES \#17 and \#18 were inserted while refining the thrust reference scenarios described in Table~\ref{tab:01}, allowing for additional sample exposure opportunity.

\section{Engineered tungsten architectures}

\subsection{Micro-castellated tungsten}\label{sec:micro-w}

Micro-castellated tungsten (\(\mu\)-W) is designed to mitigate macroscopic W cracking under recurring thermal shock through the use of disconnected near-surface W columns \cite{Terra2019,Terra2020}. Two \(\mu\)-W samples were produced by Forschungszentrum Jülich for DIII-D exposure under H-mode conditions with edge-localized modes (ELMs; hereafter ``ELMy H-mode'') [shots \#204933-\#204940]. These samples were installed in DiMES \#17 (See Table~\ref{tab:02}) and exposed to three stationary H-mode discharges (outer SP placed at R=1.47 m), followed by four rastering H-mode reference discharges. Both button specimens were tested in the 10\(^{\circ}\) angled configuration to access increased intercepted heat flux, and both were fabricated from pure W with square castellated columns of \(\sim\)690 \(\mu\)m lateral dimension separated by \(\sim\)100 \(\mu\)m gaps. They only differed in the orientation of the column cuts relative to the angled plasma-facing surface: one had castellations cut along the ``vertical'' direction, while the other had columns cut perpendicular to the inclined surface. This pair of samples was intended to probe whether castellation orientation modifies leading-edge effects and the propensity for column-scale cracking or chipping. The samples were exposed under ELMy H-mode conditions with average incident heat fluxes of \(\sim 2~\mathrm{MW\,m^{-2}}\) on flat surfaces and \(\sim 15~\mathrm{MW\,m^{-2}}\) on 10\(^{\circ}\) angled surfaces, representative of reactor-relevant first-wall and divertor loading conditions.

Following exposure, both \(\mu\)-W specimens maintained overall structural integrity and did not exhibit visible macroscopic cracking or gross material loss, indicating robust thermo-mechanical response of the castellated architecture under the enhanced H-mode plasmas with ELMs. Minor melting was limited to isolated features near the specimen periphery, predominantly affecting smaller edge columns where local geometric amplification of heat flux and reduced heat sinking is expected. Higher-magnification scanning electron microscopy (SEM) images show localized melting of an individual column [Figure~\ref{fig:02}(f)] and resolidified grains with fine intergranular microcracks [Figure~\ref{fig:02}(g)], likely formed during cooling. SEM inspection also indicates measurable surface roughening within the castellated topography, though quantitative roughness metrics are still in progress. Notably, the specimen with castellations cut perpendicular to the plasma-facing surface exhibited smaller edge-melt features than the purely vertical cut orientation. These results support continued development of micro-castellated W as a near-surface crack mitigation approach for transient heat loads, see Figure~\ref{fig:02}.

\subsection{Tungsten-fiber-reinforced composites}\label{sec:fiber-w}

Three variations of tungsten fiber-reinforced composites were exposed in DiMES holder \#17 under the same conditions described in Section~\ref{sec:micro-w}. The primary aim was to evaluate various fiber materials, lengths, orientations within a bulk W matrix, comparing their ability to arrest crack propagation and effectively increase the ductility of W materials. The materials selection builds off of prior composites recently tested in DiMES, here two samples with W fibers embedded in a W matrix (Wf/W) were prepared by Forschungszentrum Jülich \cite{Schwalenberg2022,Riesch2016}, and one sample with both W fiber and SiC fiber in a W matrix (Wf/SiCf/W) was prepared by Kyoto University. The primary Wf/W specimen was a long-fiber composite machined into a 10\(^{\circ}\) angled geometry, with fiber directions deliberately aligned both parallel and perpendicular to the plasma-facing surface to interrogate anisotropic crack-arrest behavior. Post-exposure observations (Figure~\ref{fig:02}(d,h)) show the formation of micro-cracks within the bulk W matrix; critically, these cracks are arrested/intercepted by the perpendicularly oriented W fibers, indicating effective fiber-mediated crack arrest and stress accommodation mechanisms under reactor-relevant H-mode plasmas. Crack arrest was more prevalent between vertically oriented fibers than near surface-parallel fibers, suggesting orientation-dependent stress accommodation. This behavior is consistent with prior angled-disk Wf/W exposures, though here macroscopic melt damage was avoided. Minor edge melting was observed locally on individual surface-exposed, parallel-oriented fibers. In addition to the long-fiber Wf/W, a short-fiber Wf/W button machined in a flush geometry displayed only mild surface erosion and no evidence of melting, cracking, or loss of structural integrity.

A third specimen of Wf/SiCf/W utilized a flush geometry. This sample exhibited significant localized leading-edge melting and resolidification [Figure~\ref{fig:02}(a,e)], which was not observed in prior DIII-D tests. Enhanced local heat loading, potentially associated with sample misalignment, may have contributed to this damage. The majority of the plasma-facing area remained mechanically intact with no observed failure of individual fibers, despite localized cracking in resolidified leading-edge regions. Localized surface smoothing and minor fiber-interface restructuring were observed, while bonding between constituent phases remained intact. Further analysis is required to quantify preferential erosion and/or recession of SiC relative to W in this mixed-fiber system. 

Overall, among the three composites, the long-fiber Wf/W demonstrated the clearest performance advantage: effective fiber-mediated crack arrest and no macroscopic failure under enhanced H-mode exposure with ELMs, reinforcing its promise as a plasma-facing tungsten architecture to handle transient divertor loading conditions. No major crack propagation, delamination, or structural failure was observed in any of the tungsten composite specimens. 

\subsection{Electron-beam powder-bed-fused tungsten}\label{sec:ebpbf}

\begin{figure*}[htbp]
\centering
\includegraphics[alt={EBSD orientation maps of two textured additively manufactured tungsten samples and ITER-grade tungsten.},width=0.8\linewidth]{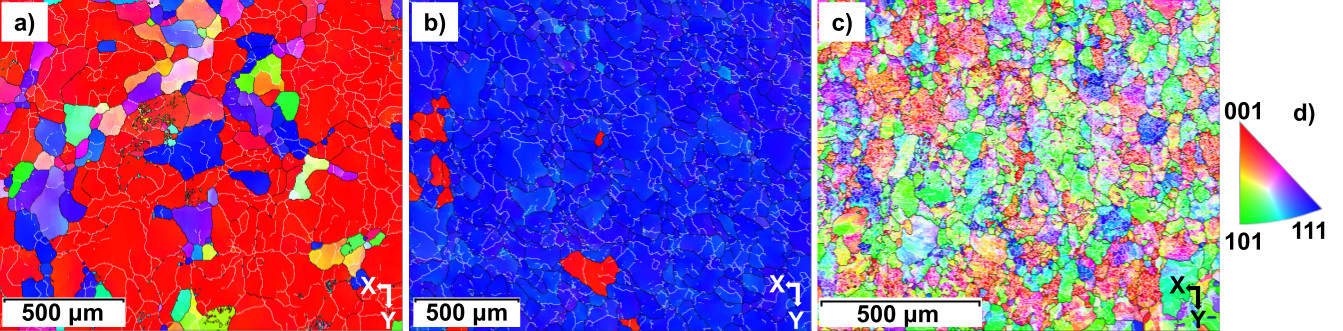}
\caption{Electron backscatter diffraction (EBSD) orientation maps of (a) predominantly (001)-textured AM-W, (b) predominantly (111)-textured AM-W, and (c) ITER-grade W, with (d) the corresponding inverse pole figure (IPF) color key. Grain boundaries are overlaid, showing low-angle grain boundaries (LAGBs, $2^\circ$--$15^\circ$, white lines) and high-angle grain boundaries (HAGBs, $>15^\circ$, black lines).}
\label{fig:03}
\end{figure*}

Electron beam powder bed fusion (EB-PBF) W samples investigated in this section (DiMES \#15--16) were produced at the Center of Additive Manufacturing and Logistics (CAMAL) at North Carolina State University (NCSU) and benchmarked against ITER-grade W produced by ALMT Japan in DIII-D under ELMy H-mode conditions.  W has a high ductile-brittle transition temperature (DBTT) (200-600 \(^{\circ}\mathrm{C}\)) \cite{Linke2019} which presents challenges during conventional processing when fabricating complex plasma facing component geometries. EB-PBF offers distinct advantages over conventional processing, including reduced oxygen content, designing intricate geometries and controlling the crystallographic textures through tuning of the electron beam parameters. EB-PBF processing produced tungsten microstructures with dominant (001) or (111) crystallographic textures, whereas the ITER-grade reference exhibited a more heterogeneous grain-orientation distribution, as revealed by electron backscatter diffraction (EBSD) analysis shown in Figure~\ref{fig:03}.

\begin{figure}[htbp]
\centering
\includegraphics[alt={SEM images of an angled EB-PBF additively manufactured tungsten sample before and after H-mode exposure.},width=\linewidth]{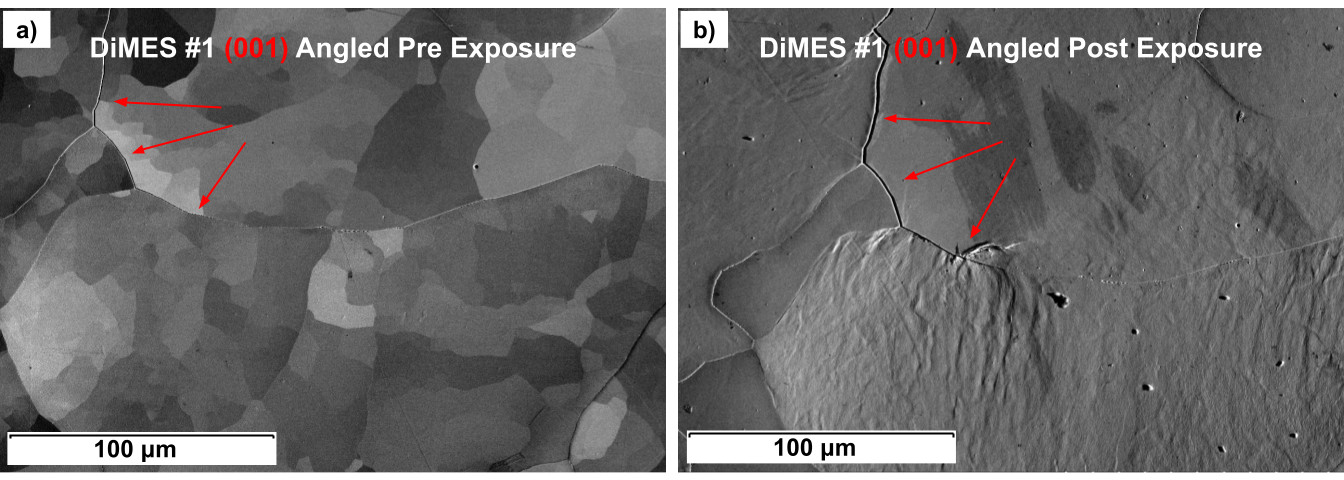}
\caption{Scanning electron microscopy (SEM) images of the angled EB-PBF AM-W sample with dominant (001) texture before and after ELMy H-mode exposure. (a) Pre-exposure surface showing pre-existing intergranular cracks (red arrows). (b) Post-exposure surface of the same region showing propagation of pre-existing cracks along grain boundaries (red arrows), accompanied by localized surface roughening and melt-like features near the crack network.}
\label{fig:04}
\end{figure}

The relative density of the EB-PBF AM-W samples exceeded 99.5\%, as determined by pycnometer measurements, despite minor process-induced porosity. As shown in Figure~\ref{fig:03}, the EB-PBF AM-W samples exhibited coarse grains with mean equivalent circle diameters (ECDs) of $73.8 \pm 41.7~\mu$m and $71.6 \pm 51.8~\mu$m for the (111)- and (001)-textured samples, respectively. In contrast, ITER-grade W exhibited a finer, equiaxed grain structure with a mean ECD of $14.7 \pm 7.3~\mu$m, approximately five times smaller than the AM-W samples. Grain boundary analysis revealed approximately equal fractions of low-angle grain boundaries (LAGBs) and high-angle grain boundaries (HAGBs) in both AM-W samples ($\sim$50\% each), whereas ITER-grade W was dominated by LAGBs (77.6\%).
Two EB-PBF-built W-angled samples varying in either dominantly (001) or (111) grain orientation were previously exposed during the first material testing campaign in Spring 2025. These samples did not show major damage to the plasma-exposed surface; therefore, this experiment focused on H-mode plasma exposures. This experiment consisted of exposing two DiMES heads with 7 samples in the same layout: 6 AM-W samples varying in geometries (10\(^{\circ}\) angled \& flat), dominant grain orientation ((001) \& (111)), build direction orientation in reference to the plasma exposed surface (\(\parallel\) \& \(\perp\)), and 1 angled reference ITER-grade W sample was included for comparison. The first DiMES head (\#15) was exposed to 6 H-mode shots  (\#204083-\#204088) with constant power input of 5.9 MW while the second DiMES head (\#16) experienced 7 H-mode shots  (\#204089-\#204095) with an increase of NBI power input of 0.5 MW after every shot reaching up to 7.1 MW. Following plasma exposure, the samples did not show any visual damage to the top surface. Figure~\ref{fig:04}, shows scanning electron microscopy before and after plasma exposure of the (001) AM-W angled sample from the first DiMES head. Intergranular cracks were observed in the AM-W samples prior to exposure in DIII-D, which may be associated with oxygen segregation at grain boundaries during solidification \cite{Ellis2021}. After plasma exposure, localized melt-like features and surface roughening were observed near pre-existing intergranular cracks.

The angled AM-W and ITER-grade W samples experienced higher heat flux due to the top surface angled geometry of 10\(^{\circ}\). The first DiMES head experienced heat fluxes corresponding to the reference rastering H-mode in Table~\ref{tab:01}. With additional NBI heating, the second DiMES head experienced slightly higher heat loads of \(2.2~\mathrm{MW\,m^{-2}}\) on flat samples and \(\sim 17~\mathrm{MW\,m^{-2}}\) on the angled samples. Despite the increased thermal loading, no macroscopic damage, cracking, or material loss was observed, and the AM-W samples behaved similarly to the ITER-grade reference. Localized melt-like features and roughening were confined to regions near pre-existing intergranular cracks, indicating that manufacturing-induced defects dominated the surface evolution under these conditions. These results support EB-PBF tungsten as a viable manufacturing route for plasma-facing components.

\subsection{Laser powder-bed-fused tungsten}\label{sec:lppbf}

Conventionally manufactured (CM) and additively-manufactured pure tungsten specimens were exposed to L- and H- Mode plasmas in DiMES. CM-W samples of pure (99.95\%) tungsten were sourced from Midwest Tungsten in the standard DiMES 6 mm diameter top hat geometry. AM-W samples were manufactured by the Alloy Design \& Development (ADD) Laboratory at the University of Wisconsin--Madison using laser powder bed fusion on an EOS M290 system. The AM-W specimens were processed with a volumetric energy density of 602 J mm\(^{-3}\) to achieve a mass density of 18.7 g/cm\(^{3}\) (\(\sim\)97\% theoretical density) with low porosity and minimal microcracking. Both types of specimens were polished to a mirror finish, down to 1 \(\mu\)m diamond suspension, before pre-exposure characterization using scanning electron microscopy (SEM), energy dispersive X-ray spectroscopy (EDS), and laser scanning confocal microscopy (LSCM).

\begin{figure*}[!t]
\centering
\includegraphics[alt={Panel grid of surface microscopy and compositional measurements before and after L-mode and H-mode exposure.},width=0.76\linewidth]{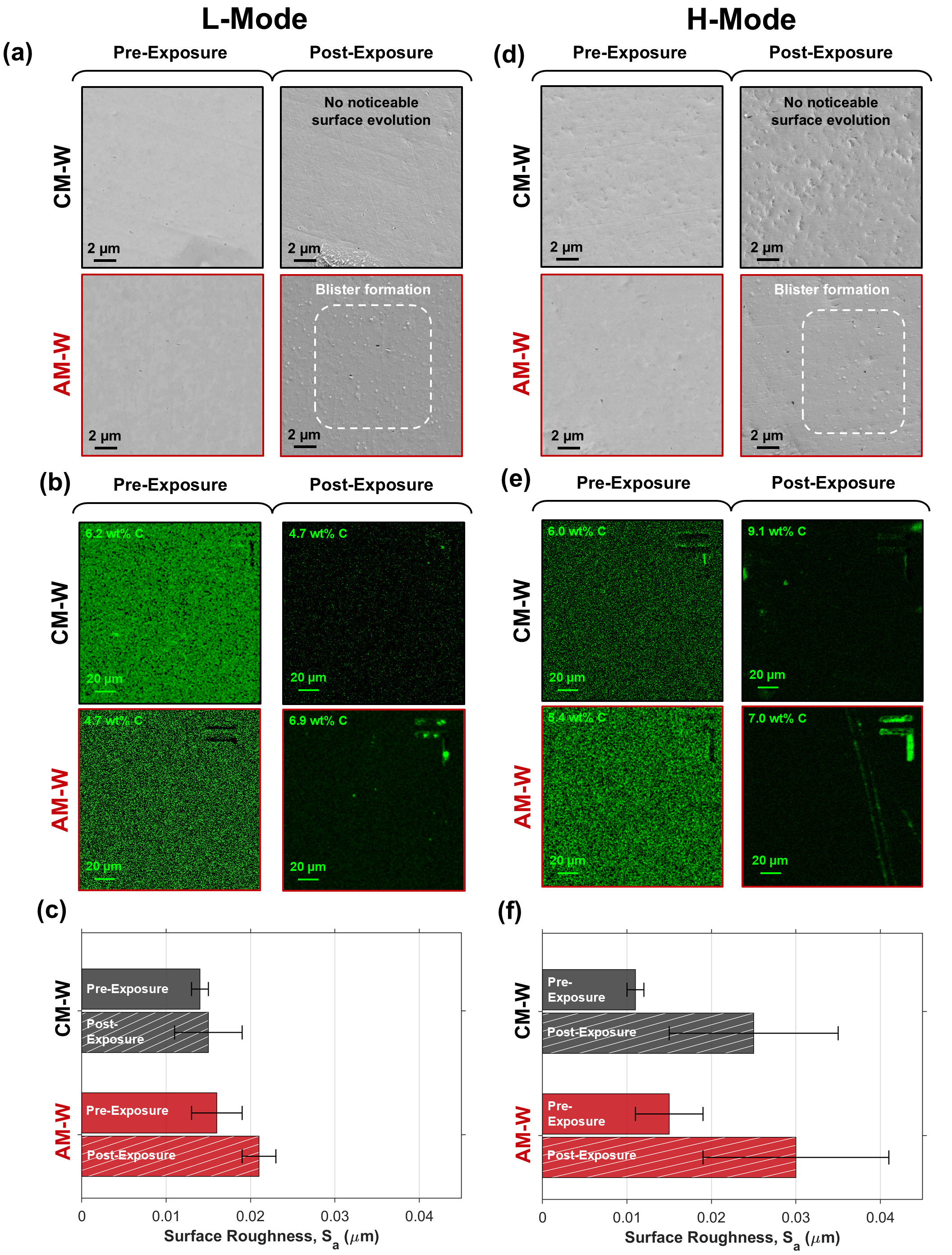}
\caption{Pre- and post-exposure characterization of laser powder bed fusion additively manufactured tungsten (AM-W) and conventionally manufactured tungsten (CM-W) exposed to L-mode (a--c) and H-mode (d--f) plasmas. (a,d) Scanning electron microscopy (SEM) images show limited overall surface evolution following exposure for all specimens, with localized blister-like features on AM-W highlighted by dashed outlines. (b,e) Energy-dispersive X-ray spectroscopy (EDS) maps showing the spatial distribution of carbon-associated signal. Annotated carbon contents are apparent EDS weight percentages (wt.\%) calculated in AZtec with C, O, and W selected and built-in spectral deconvolution enabled. Overlap between C~K$\alpha$ and W N-series emission limits their quantitative interpretation; the carbon signal may include contributions from deposits, dust, and pre-existing surface contamination. (c,f) Areal surface roughness, expressed as the arithmetic mean height, $S_a$, calculated from laser scanning confocal microscopy (LSCM) topography maps following noise removal and tilt correction. Surface roughness increased for both CM-W and AM-W following H-mode exposure. Bars represent mean values, and error bars indicate $\pm$ one standard deviation across measurement locations.}
\label{fig:05}
\end{figure*}

CM-W and AM-W specimens were exposed simultaneously, with L-Mode specimens subject to five plasma shots (\#203730-203734), with a time-averaged perpendicular heat flux of \(1.2~\mathrm{MW\,m^{-2}}\). H-mode specimens were subject to six reference plasma shots (\#203736-203741), with an inter-ELM heat flux of \(1.5~\mathrm{MW\,m^{-2}}\) and intra-ELM heat fluxes of \(3.7~\mathrm{MW\,m^{-2}}\) at ELM frequencies of 44~Hz. In H-mode, the line-averaged electron densities were approximately \(8\times10^{12}~\mathrm{cm^{-3}}\) with electron temperatures up to 50 eV, and OSP rastering was employed. Specimens were characterized post-exposure using SEM, EDS, and LSCM.
SEM was used to compare surface morphology before and after plasma exposure, with focused-ion-beam (FIB) fiducial marks used to relocate corresponding surface regions. EDS quantification for the maps in Figure~\ref{fig:05} was performed in AZtec, with the quantification element list restricted to C, O, and W and the built-in spectral deconvolution option enabled to calculate elemental weight percentages. Spectra used for the subsequent comparisons were exported after initial software processing in AZtec, except for the L-mode AM-W pre-exposure spectrum, which was acquired and processed using a Thermo Fisher Scientific EDS system with Pathfinder software on a Zeiss Gemini 450. The reported carbon contents are apparent compositions conditional on the selected elements and spectral model; deconvolution does not by itself establish that the overlapping C~K$\alpha$ and W N-series contributions are accurately separated. LSCM topography maps were processed using noise removal and tilt correction before the software calculated the areal arithmetic mean height, $S_a$. Figure~\ref{fig:05}(c,f) reports mean roughness values across measurement locations, with error bars representing $\pm$ one standard deviation.

Cold-sprayed refractory Ta and Ta-W coatings were also exposed under matched DiMES L- and H-mode scenarios for comparison with LPBF AM-W and CM-W. Prior Ta CS laboratory studies showed \(\sim\)3.5\(\times\) higher D retention than polycrystalline Ta and \(\sim\)100\(\times\) higher retention than W \cite{Ialovega2025}. The DIII-D post-exposure analysis will be reported separately.

Post-exposure SEM imaging of L-Mode specimens, shown in Figure~\ref{fig:05} (a), revealed no significant surface evolution in CM-W, whereas possible deuterium blistering was observed on AM-W. Further analysis of these features is ongoing. Post-exposure EDS maps in Figure~\ref{fig:05} (b) show carbon-associated signal on the specimen surfaces, including enhanced signal in the L-shaped FIB fiducial marks. These observations are consistent with carbon-bearing surface deposits. Erosion of the graphite DiMES holder and carbon-based first wall is a plausible source, although EDS alone cannot distinguish plasma-deposited carbon from carbon-bearing dust or pre-existing surface contamination. Figure~\ref{fig:05} (c) shows a slight increase in surface roughness for AM-W specimens, following L-Mode exposure, while the surface roughness for CM-W is unchanged within the measurement uncertainty of \(\pm\) 1 standard deviation. Further characterization will include X-ray Diffraction (XRD), X-ray photoelectron spectroscopy (XPS), and thermal desorption spectroscopy (TDS), to determine the deuterium retention in the specimens.

Similar trends to the L-Mode exposures were observed in H-Mode specimens. Post-exposure SEM images in Figure~\ref{fig:05} (d) show no pronounced morphological changes in CM-W at the imaged length scales, while localized blister-like features are visible on AM-W. Post-exposure EDS maps in Figure~\ref{fig:05} (e) show carbon-associated signal on both CM-W and AM-W. Comparison of the corresponding pre- and post-exposure spectra shows an increased low-energy feature near the carbon emission energy relative to the W M-peak region, supporting a qualitative interpretation of an additional carbon contribution after H-mode exposure. In contrast to L-Mode, the H-Mode exposures resulted in a notable increase in surface roughness, with both specimens exhibiting an approximate +0.015 \(\mu\)m increase, a factor of 2, as shown in Figure~\ref{fig:05} (f). The increase in surface roughness could have implications for the sputtering and erosion of plasma-facing materials and may impact infrared (IR) thermography measurements of the surface temperature.

In summary, both CM-W and AM-W specimens exhibit minimal surface evolution in L-Mode and H-Mode plasmas. No macroscopic cracking, melting, or material loss was observed in either CM-W or AM-W under the tested conditions. The broadly comparable response of both specimen types supports further evaluation of LPBF tungsten for first-wall components. Future work will include ion irradiation of specimens, to degrade the thermal conductivity, and angled specimen geometries, to increase the impingent heat and particle fluxes.

\section{Modified tungsten compositions and defect states}

\subsection{Potassium-doped tungsten}\label{sec:k-w}

K-doped-W (EP-1) produced by ALMT Japan was benchmarked against ITER W also produced by ALMT. A series of standard flat and 10\(^{\circ}\) angled DiMES \cite{Rudakov2017} K-doped-W and ITER-W samples were exposed within DIII-D to the reference attached ELMing H-mode plasmas in DiMES \#3 (shots \#203742-203748), and DiMES \#8 (shot \#203769).

\begin{figure}[htbp]
\centering
\includegraphics[alt={Photos of a DiMES sample holder containing tungsten, K-doped tungsten and tungsten-rhenium before and after H-mode exposure.},width=0.99\linewidth]{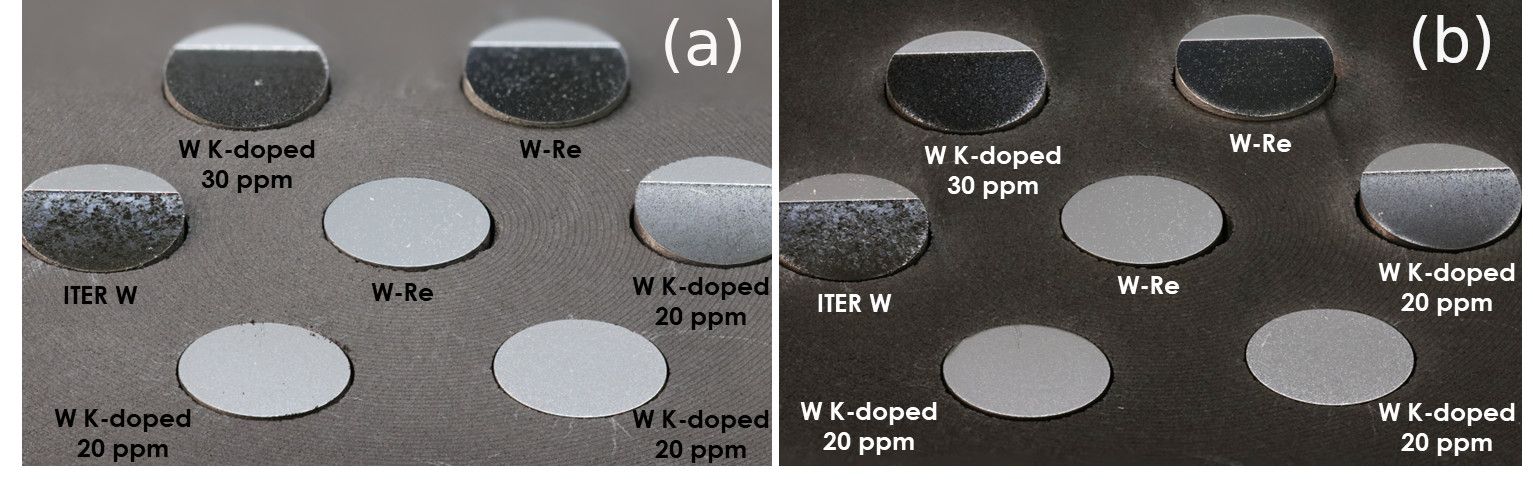}
\caption{Pre-exposure (a) and post-exposure (b) photographs of DiMES holder \#3 containing ITER-grade W, K-doped W, and W-Re alloy samples after rastering H-mode plasma exposure.}
\label{fig:06}
\end{figure}

K-doped W approached the ITER W reference in mass loss performance shown in Figure~\ref{fig:07} and showed comparatively limited visible surface morphology evolution in optical microscopy, shown in Figure~\ref{fig:06}. This is encouraging because K-doped W is being developed to improve high-temperature stability and recrystallization resistance in tungsten-based plasma-facing materials \cite{Nogami2024,Nogami2020}.

In a separate low-power ohmic exposure set using the 5\(^{\circ}\) angled HFIR-compatible DiMES holder (DiMES \#13), K-doped W was exposed alongside pristine ITER-grade W, neutron-irradiated W, and TiB\(_{2}\) for 16 discharges to compare D uptake and desorption behavior under matched conditions. Following 16 Ohmic-heated discharges (approximately 32~s nominal cumulative DiMES dwell time), the specimens were analyzed by thermal desorption spectroscopy. Both deuterium uptake and desorption temperatures in the K-doped W sample were comparable to pure ITER-grade W. Total desorbed D\(_{2}\) was 2.88 \(\times\) 10\(^{-7}\) mol for K-doped W and 2.05 \(\times\) 10\(^{-7}\) mol for ITER-grade W, corresponding to \(\approx\)1.4\(\times\) higher D\(_{2}\) release from K-doped W, with peak desorption temperatures of 704 \(^{\circ}\)C and 710 \(^{\circ}\)C, respectively. The increase in D\(_{2}\) uptake can be attributed to greater intrinsic defects in the K-doped W due to greater numbers of grain boundaries and is consistent with previous linear plasma device exposures comparing uptake between K-doped W and pure W \cite{Ma2023}. Because K-doped W and neutron-irradiated W showed different desorption behavior despite similar deposition levels in this holder, the K-doped W retention response was not interpreted as being dominated by impurity deposition. Overall, K-doped W exhibited D\(_{2}\) desorption temperatures comparable to pristine ITER-grade W but a moderately higher total D\(_{2}\) retention (\(\approx\)1.4\(\times\)), whereas the neutron-irradiated W reference showed a much larger retention increase, as discussed in Section 4.4 (Figure~\ref{fig:09} and \ref{fig:10}).

\subsection{Additively manufactured W--10Ta}\label{sec:w-ta}

Additively manufactured (AM) W-Ta, 10\% Ta by weight, produced through laser powder bed fusion (LPBF), by the Metallurgy and Materials department at the University of Birmingham UK, was benchmarked against ITER W produced by ALMT Japan. The measured density of the W--Ta material was 18.42~g\,cm$^{-3}$. Flush W--Ta and angled LPBF W--10Ta specimens were exposed alongside ITER-W references in DiMES holders \#7--8 under rastered ELMy H-mode conditions (Table~\ref{tab:02}), including shots \#203759--\#203765.

\begin{figure}[!t]
\centering
\includegraphics[alt={Bar chart of tungsten alloy mass losses at two exposure heat fluxes.},width=0.8\linewidth]{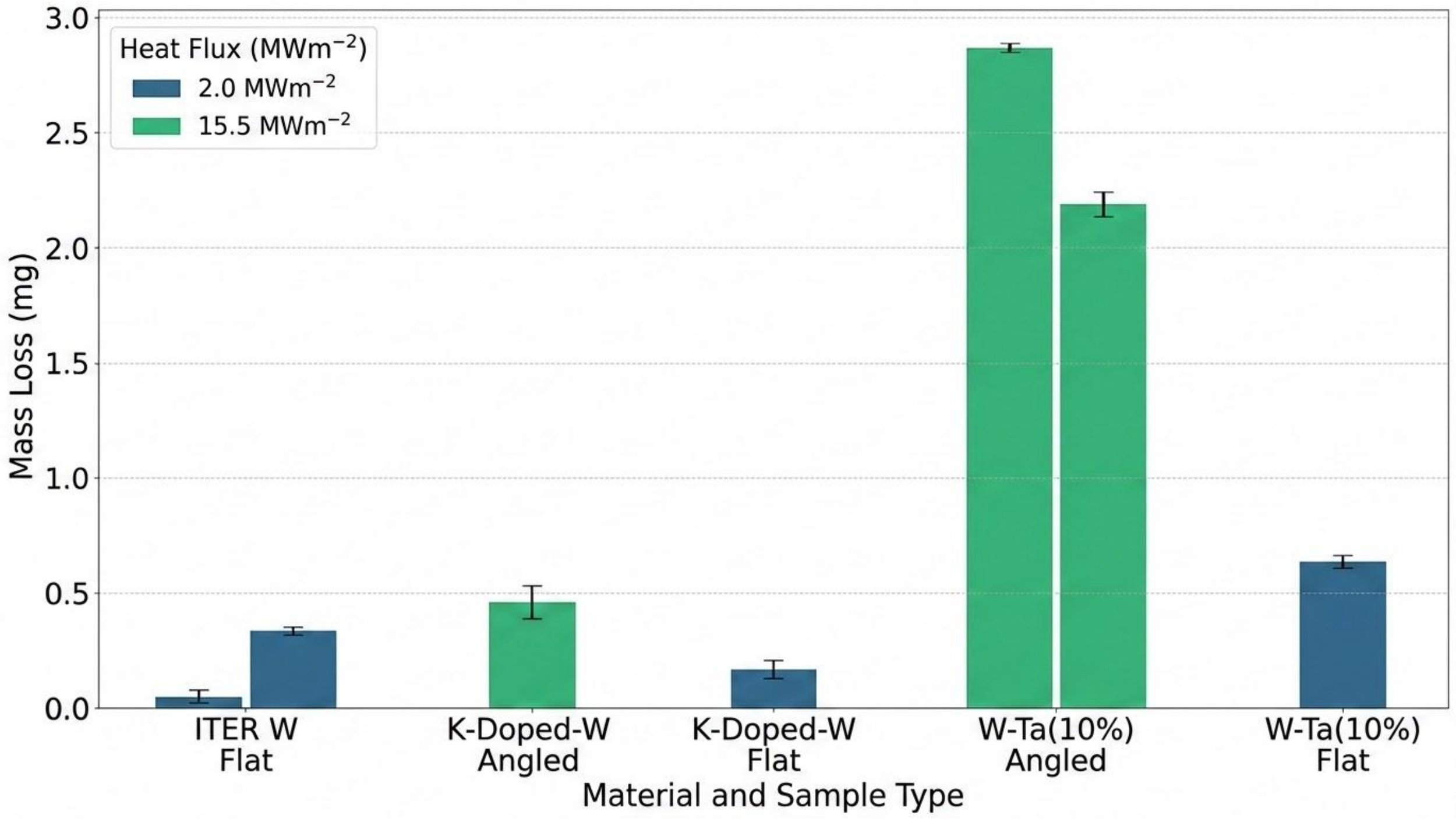}
\caption{Post-exposure mass loss by individual sample, grouped by material, sample geometry, and nominal incident heat flux. Blue bars correspond to flat samples exposed at \(\sim\)2.0 MW m\(^{-2}\), and green bars correspond to 10\(^{\circ}\) angled samples exposed at \(\sim\)15.5 MW m\(^{-2}\). Error bars indicate within-sample measurement uncertainty from repeated pre- and post-exposure weighings.}
\label{fig:07}
\end{figure}

The mass loss comparison in Figure~\ref{fig:07} provides the clearest first order performance discriminator. The two flat ITER W reference samples lost 0.0500 mg and 0.3350 mg,  a spread that exceeds the within sample measurement uncertainty and is large enough that the ITER W samples cannot be treated as a single, precise erosion baseline. Despite this scatter, the flat W-Ta sample showed clearly resolved mass loss above both ITER-W values, approximately twice the higher ITER-W reference value. The 10\(^{\circ}\) angled W--Ta samples showed a larger response, with mass loss \(\sim\)3--5\(\times\) higher than the flat W-Ta sample. The elevated W--Ta mass loss and its amplification under angled exposure are robust findings, while the precise W--Ta/ITER-W ratio remains subject to further cross-calibration.

\begin{figure*}[htbp]
\centering
\includegraphics[alt={Three SEM views of plasma-exposed W-10Ta surfaces showing texture and localized material damage.},width=0.79\linewidth]{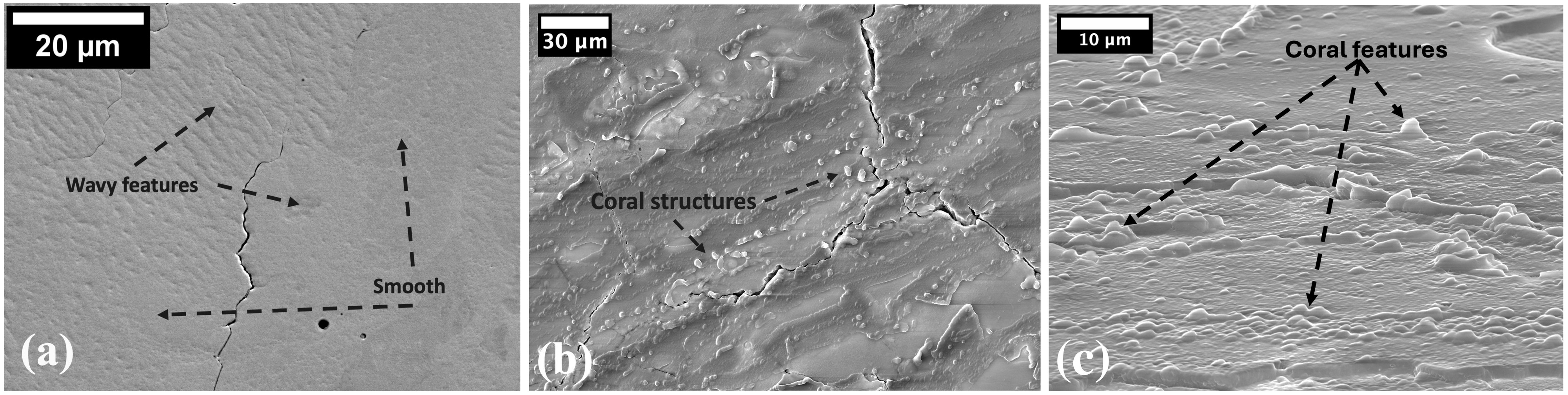}
\caption{SEM images of plasma-exposed W-10Ta surfaces. (a) Flat sample showing adjacent wavy and smooth regions, suggesting a grain-dependent sputtering response. (b) 10\(^{\circ}\) angled sample showing enhanced surface roughening and bright coral-like features. (c) Additional SEM image of the angled sample at 70\(^{\circ}\) tilt, revealing the raised topography of these features.}
\label{fig:08}
\end{figure*}

SEM imaging of the exposed flat W-Ta sample, in Figure~\ref{fig:08}(a), shows spatially distinct surface morphologies, including wavy regions adjacent to smoother regions. The localization of these features suggests a grain-orientation-dependent sputtering response. In contrast, the angled W-Ta samples showed a rougher morphological response to the greater incident heat fluxes, with 'coral'-like structures forming on the surface shown in Figure~\ref{fig:08}(b). The bright coral-like features are associated with raised surface structures, as shown by the 70\(^{\circ}\) tilted SEM image in Figure~\ref{fig:08}(c). Their origin remains uncertain; possible contributions include carbon deposition and localized surface modification under enhanced thermal loading. Reduced thermal diffusivity, previously reported for W–Ta alloys \cite{Nogami2022}, may also contribute. Further investigation of the formation mechanisms is ongoing and will be reported elsewhere.

\subsection{Tungsten alloys: W-Re and W-Ti-Cr}\label{sec:w-alloys}

In addition to the aforementioned W materials, two types of W alloys were tested under the reference rastering H-mode conditions. The first is a commercial W-Re alloy composed of 10 wt\% Re, supplied by ALMT. W-Re alloys offer reduced DBTT and improved ductility relative to the undoped ITER-grade W reference used in this study, while K-doped W is primarily intended to improve high-temperature stability and recrystallization resistance. The other is a nanostructured W-Ti-Cr alloy developed by Stony Brook University that combines nanostructuring (nm-grain sizes) of Cr precipitates with Ti doping of W grain boundaries. These materials are synthesized via a carefully controlled powder metallurgy and sintering process, resulting in nanostructured alloys designed to achieve enhanced thermal stability and radiation tolerance \cite{Donaldson2018,Olynik2022}. Commercial W--10Re and K-doped W were tested alongside ITER-W in DiMES \#3, whereas W--Ti--Cr and ORNL EB-PBF W--Re were included in DiMES \#4 (Table~\ref{tab:02}). Commercial W-Re and K-doped W specimens were exposed during seven H-mode discharges, \#203742--203748, together with ITER-W reference samples. The nanostructured W-Ti-Cr alloy was exposed separately during shots \#203749--203755, together with ORNL EB-PBF W, EB-PBF W--Re, and UHTC specimens. The samples were exposed to 6--7 H-mode discharges with an average steady-state perpendicular heat flux of \(\sim\)2.4 MW m\(^{-2}\), ELM heat fluxes of \(\sim\)6 MW m\(^{-2}\), and ELM frequencies of \(\sim\)75 Hz; 10\(^{\circ}\) angled specimens intercepted heat fluxes above 10 MW m\(^{-2}\). Thermocouples indicated peak back-surface temperatures of \(\sim\)710--845 \(^{\circ}\)C across shots \#203742--203748. In general, the flush-mounted samples exhibited benign macroscopic response. Optical and SEM inspection showed no major roughening or cracking, with only some enhanced erosion on the flush W-Ti-Cr specimen due to a slightly raised leading edge; the central location remained nearly pristine. Overall, the commercial W-Re and K-doped W alloys showed degradation modes comparable to the ITER-grade W reference, without new alloy-specific failure modes.

\begin{figure}[htbp]
\centering
\includegraphics[alt={SEM images of the angled W-Ti-Cr sample showing extensive cracks and detailed crack morphology after exposure.},width=0.99\linewidth]{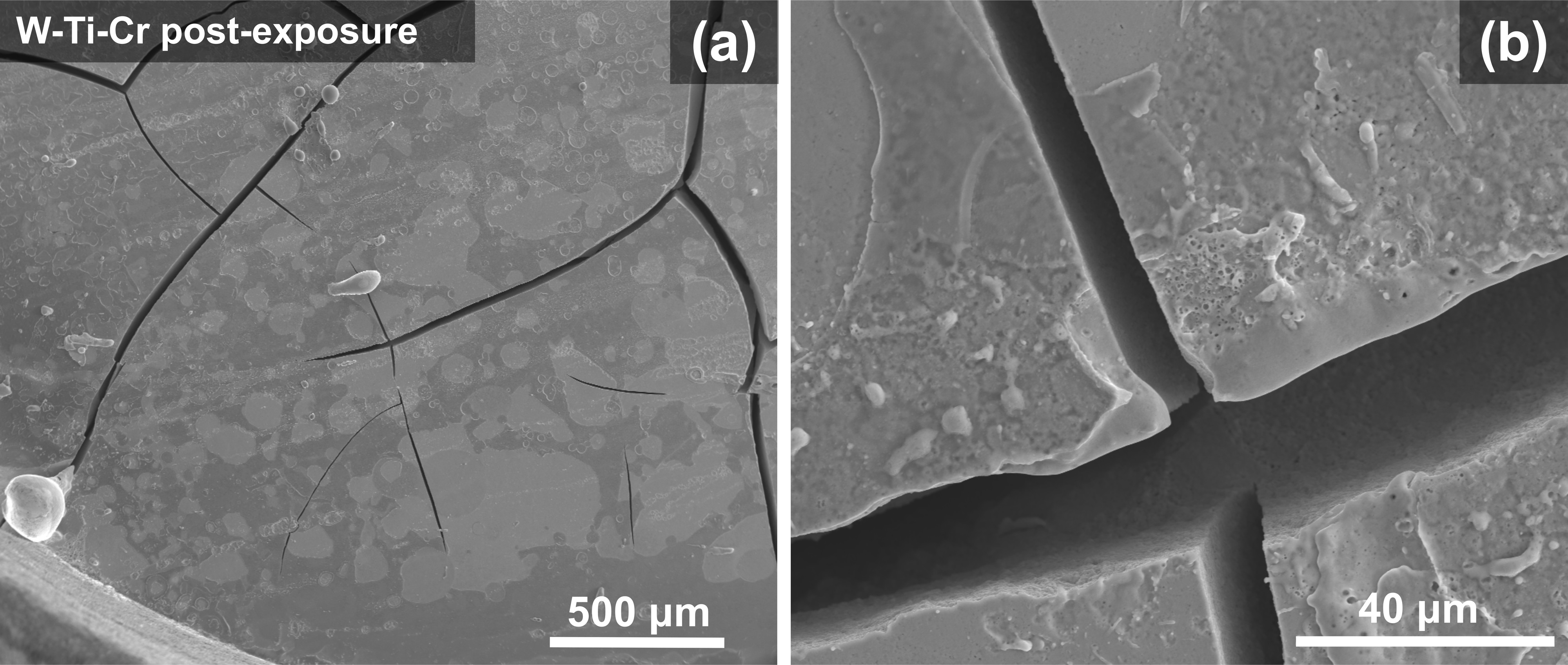}
\caption{Post-exposure SEM images of the angled W--Ti--Cr specimen following rastered H-mode plasma exposure. (a) Overview showing an extensive network of surface cracks. (b) Higher-magnification image showing intersecting cracks with widths of approximately 10--20~\(\mu\)m and localized surface modification near the crack edges.}
\label{fig:08a}
\end{figure}

The 10\(^{\circ}\) angled specimens displayed differing levels of surface integrity among the W alloys. The angled W-Re sample showed a surface response comparable to the ITER-W and K-doped W, with no optically visible melting, cracking, or unexpected degradation (see Figure~\ref{fig:06} (b)). SEM reveals micron-scale surface roughening within individual, preferentially-oriented grains, similar to that observed for the doped and AM materials in Sections \ref{sec:ebpbf}, \ref{sec:lppbf}
and \ref{sec:k-w}. This morphology can be described as ``grain boundary grooving'', which occurs at elevated surface temperatures as a result of competition between grain growth and surface diffusion effects \cite{Haremski2022}. Similar morphology was observed on angled W specimens exposed at surface temperatures above \(\sim\)2000 \(^{\circ}\)C, consistent with the onset of recrystallization-related surface evolution. Quantification of surface roughness via Ellipsometry shows similar levels of roughness increase between ITER-W, K-doped W, and W-Re samples exposed on DiMES \#3 \&4. In contrast, the angled nanostructured W-Ti-Cr alloy displayed deep surface cracking and localized melting in the `hot spot' region of the angled samples. The sample surface developed an extensive crack network, with crack widths of approximately 10--20~\(\mu\)m and localized melting near larger surface discontinuities (Figure~\ref{fig:08a}). A discrete melted/eroded patch (\(\sim\)400 x 800 \(\mu\)m) was correlated with an anomalous W-emission burst observed with in-situ diagnostics (DiMES TV/MDS/core spectroscopy). Notably, these deep surface cracks did not result in specimen fragmentation; the entire angled specimen was successfully removed as one piece for post-exposure analysis. Auger spectroscopic mapping indicated elevated oxygen and reduced Ti content in the melted regions, while Ti remained detectable in the undamaged near-surface at the level of a few atomic percent (\(\sim\)5 at\%). Two potential explanations for the material failure, irrespective of the alloying concept, are noted. Post-mortem analysis revealed that this particular sample was polished out of spec, resulting in a steeper inclination angle (\(\sim\)14\(^{\circ}\)) and consequently higher perpendicular heat flux. Additionally, large disparities in the thermocouple signals suggest a lower than expected thermal conductivity or poor thermal contact. These results indicate that W-Re and K-doped W retained ITER-W-like PMI behavior under the tested H-mode conditions, while the W-Ti-Cr failure was likely amplified by off-nominal angle and thermal-contact effects; follow-up W-Ti-Cr exposures should use controlled 10\(^{\circ}\) geometry, verified thermal contact, and matched ITER-W/K-doped W references.

\subsection{Neutron-irradiated ITER-grade tungsten}\label{sec:irr-w}

\begin{figure*}[htbp]
\centering
\includegraphics[alt={Micrograph matrix of pristine tungsten, neutron-irradiated tungsten, titanium diboride, and K-doped tungsten before and after exposure.},width=0.7\linewidth]{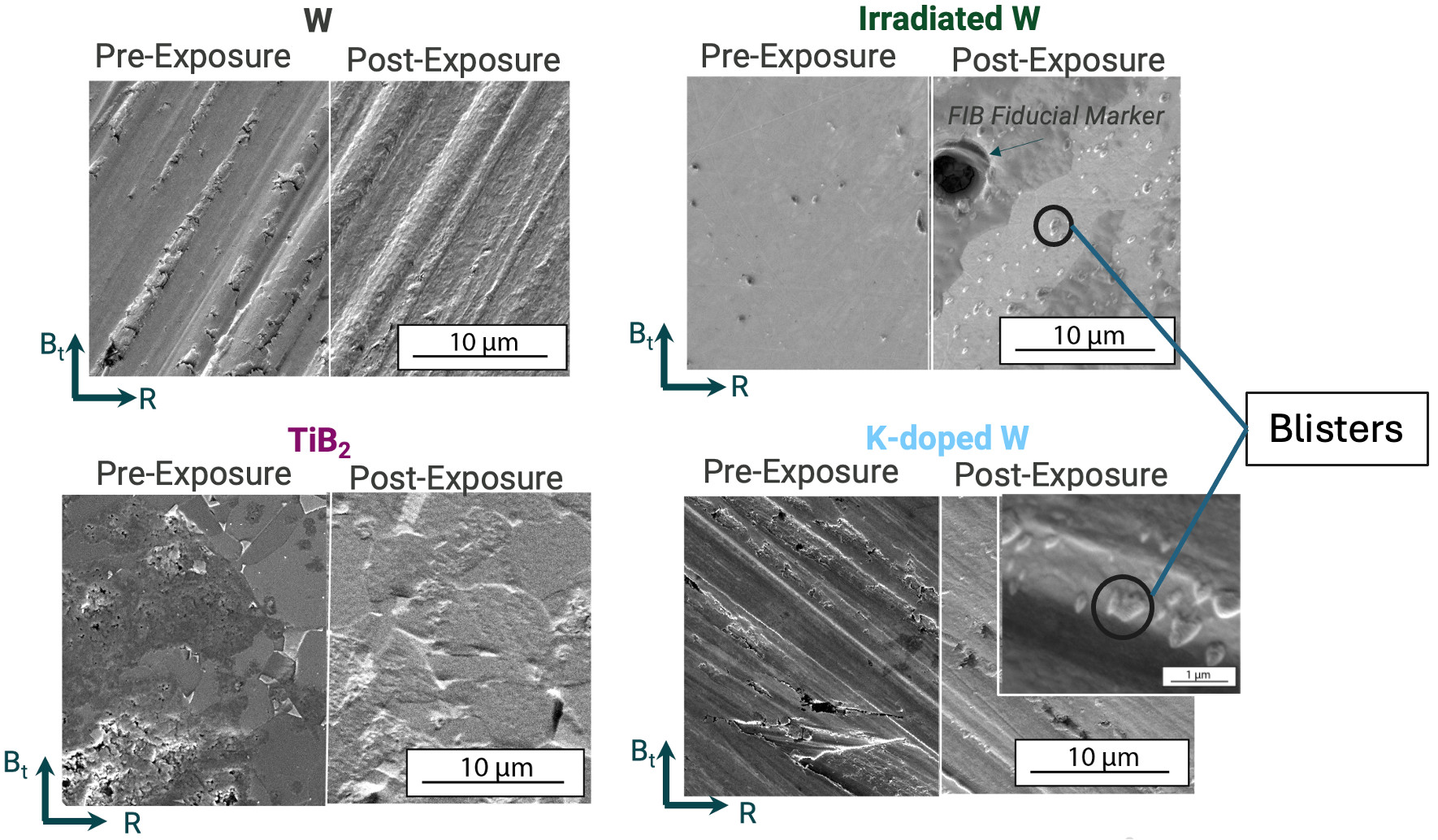}
\caption{Pre- and post-exposure SEM images of pristine W, neutron-irradiated W (0.3 dpa at 550 \(^{\circ}\)C), K-doped W, and TiB\(_{2}\) samples after 16 ohmic-heated DIII-D plasma exposures. Surface roughening is visible on W and TiB\(_{2}\). Blister-like features are more apparent on the outboard W-based samples, including neutron-irradiated W and K-doped W, where impurity deposition was also greater. Cross-sectional analysis also indicates larger blister features (\textasciitilde{}tens of $\mu$m scale) on ITER-grade, non-irradiated W.}
\label{fig:09}
\end{figure*}

\begin{figure}[htbp]
\centering
\includegraphics[alt={Plots of thermal desorption rates and cumulative deuterium release for irradiated and reference materials.},width=0.99\linewidth]{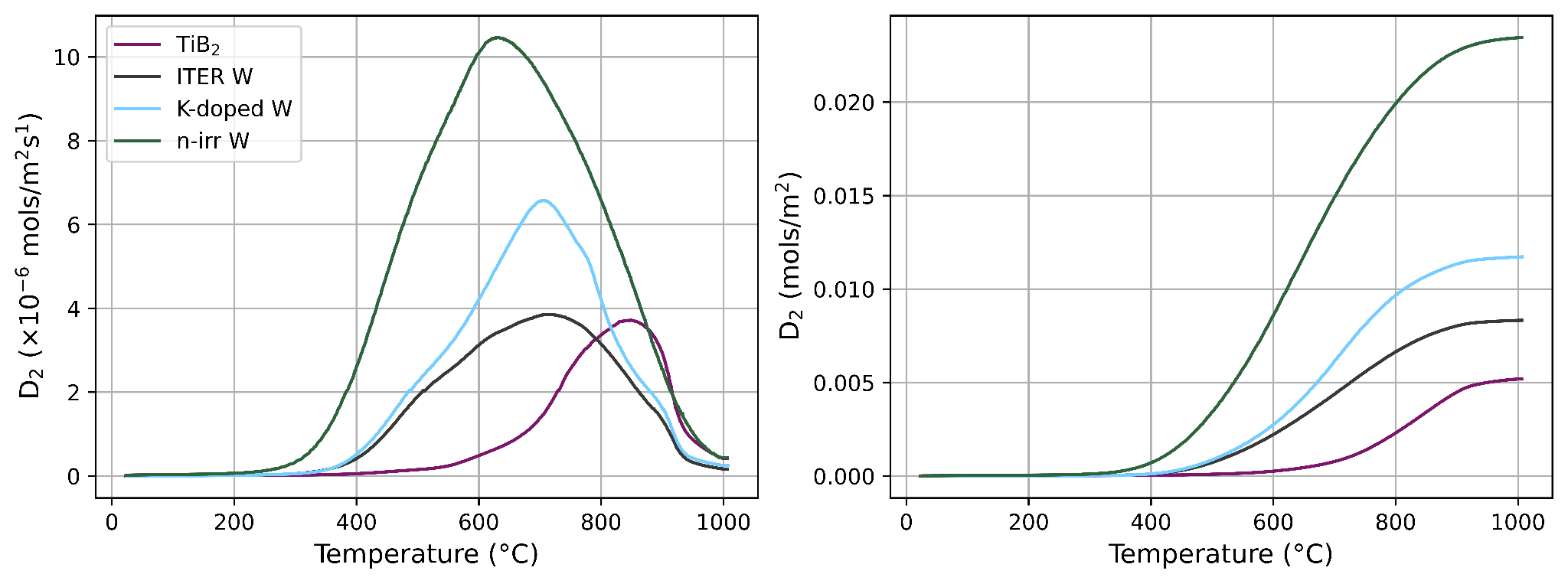}
\caption{Thermal desorption spectroscopy measurements of D\(_{2}\) release from plasma-exposed pristine W, neutron-irradiated W, K-doped W, and TiB\(_{2}\) samples following ohmic-heated DIII-D exposure. The left plot is the D\(_{2}\) flux as a function of temperature. The right plot shows cumulative D$_2$ release as a function of temperature.}
\label{fig:10}
\end{figure}

In the first integrated DIII-D divertor exposure of neutron-irradiated plasma-facing materials, HFIR-irradiated ITER-grade W, pristine ITER-grade W, K-doped W, and TiB\(_{2}\) were exposed in a newly designed angled DiMES holder during 16 ohmic-heated discharges (within shot range \#203789--203804) to build deuterium fluence while keeping sample temperatures below the 550 \(^{\circ}\)C irradiation temperature, to avoid annealing neutron-induced defects. The holder incorporated a TZM retaining plate to clamp the thin 0.5 mm HFIR specimens and a 5\(^{\circ}\) inclined geometry, yielding 0.8--1.35 MW m\(^{-2}\) peak heat fluxes on the samples (compared with 0.2--0.3 MW m\(^{-2}\) on flush divertor tiles) while avoiding melting and limiting the plasma-shadowed fraction to \(\sim\)13 \% of each button surface. The neutron-irradiated W had been exposed in HFIR to \(\sim\)0.3 dpa at 550 \(^{\circ}\)C, corresponding to a few weeks of operation at DEMO/FPP-relevant neutron fluxes, before DIII-D exposure. A geometry-corrected estimate based on the reference peak field-aligned particle flux, a $6.6^\circ$ incidence angle, and $\sim32$~s nominal dwell gives a cumulative D$^+$ fluence of order $2\times10^{24}$~m$^{-2}$ on the unshadowed inclined surface; shot-to-shot flux variations are not included. Compared with pristine W, the neutron-irradiated W exhibited 2.8\(\times\) higher total D\(_{2}\) uptake, increasing from 2.05 \(\times\) 10\(^{-7}\) mol for pristine ITER-W to 5.77 \(\times\) 10\(^{-7}\) mol for HFIR-W, consistent with an increased concentration of deuterium trapping sites introduced by neutron damage. This effect is well documented but has not been thoroughly demonstrated under tokamak-plasma conditions with neutron degradation \cite{Schwarz-Selinger2023}. Figure~\ref{fig:09} compares pre- and post-exposure surface morphologies of pristine W, neutron-irradiated W, K-doped W, and TiB\(_{2}\), while Figure~\ref{fig:10} shows the corresponding D\(_{2}\) TDS spectra and integrated release. The dominant D\(_{2}\) desorption peak for neutron-irradiated W occurs at \(\sim\)630 \(^{\circ}\)C, \(\sim\)80 \(^{\circ}\)C lower than the \(\sim\)710 \(^{\circ}\)C peak for pristine W, indicating a shift toward lower binding-energy traps consistent with irradiation-induced vacancy-type point defects and defect clusters. Although impurity deposition (predominantly C/O-rich layers) varied across the DiMES head, the lack of deposition-related TDS peaks and the differing desorption behavior of similarly deposited neutron-irradiated W and K-doped W suggest that neutron-irradiation damage, not deposition, primarily drove the increased D\(_{2}\) retention.

Post-exposure SEM indicated that surface morphology was also strongly affected by radial position and impurity deposition: roughening was more prominent on the reference W and TiB$_2$ located at slightly lower $R$ closer to the strike point, whereas blistering was more evident on the neutron-irradiated W and K-doped W located at higher $R$ (see Figure~\ref{fig:01}). Cross-sectional analysis showed the largest individual blisters on pristine ITER-grade W.

The results constitute, to the authors' knowledge, the first quantitative demonstration of neutron-damage-driven retention increase in a tokamak divertor plasma, extending prior ion-beam-based studies to a realistic PMI and neutron degradation environment.

\section{Alternative refractory metallic materials}

\subsection{Refractory multi-principal-element alloys}\label{sec:mpea}

Refractory multi-principal-element alloys (MPEAs) were evaluated as candidate plasma-facing and structural materials for compact fusion systems. For Avalanche, refractory MPEAs provide a tunable alloy design space for high-temperature, radiation-tolerant components in a compact electrostatic fusion architecture, including potential structural and radiation-conversion applications. Recent results in quaternary and quinary metallic high-entropy alloys (HEAs) showed promising irradiation and He resistance properties with physical damage and swelling being negligible in quinary W based systems \cite{ElAtwani2019}\cite{ElAtwani2023}. The refractory MPEAs/HEAs were designed using an AI/ML-assisted "predict-first" workflow combining first-principles density functional theory (DFT) calculations with thermodynamic screening metrics, including entropy/enthalpy of mixing, atomic size mismatch, and melting point, to identify refractory alloys with targeted single-phase body-centered-cubic (bcc) stability for plasma-facing fusion applications. \cite{ElAtwani2023,Cantor2024,Divilov2024,Miracle2017}. A relatively simple approach was used to first screen alloy compositions by looking at the interplay between configurational entropy and the enthalpy of mixing in the crystal phase stability of the MPEAs. From a thermodynamic perspective, alloy design focused on maximizing configurational entropy while maintaining near-zero to moderately negative mixing enthalpies, avoiding ordered intermetallics at (\(\Delta H_{\mathrm{mix}} \ll 0\)) and phase separation at (\(\Delta H_{\mathrm{mix}} \gg 0\)). Furthermore, to obtain trends in alloy solid solutions, DFT based approaches were used to approximate the predicted crystal phase when low amounts ( < 10 at\%) of bcc stabilizer elements were used in largely non-BCC systems. However, atomic size differences in such systems can introduce significant lattice strain and undesirable mechanical properties.

\begin{figure*}[htbp]
\centering
\includegraphics[alt={Surface photograph and color microscopy maps of a refractory multi-principal element alloy sample set.},width=0.7\linewidth]{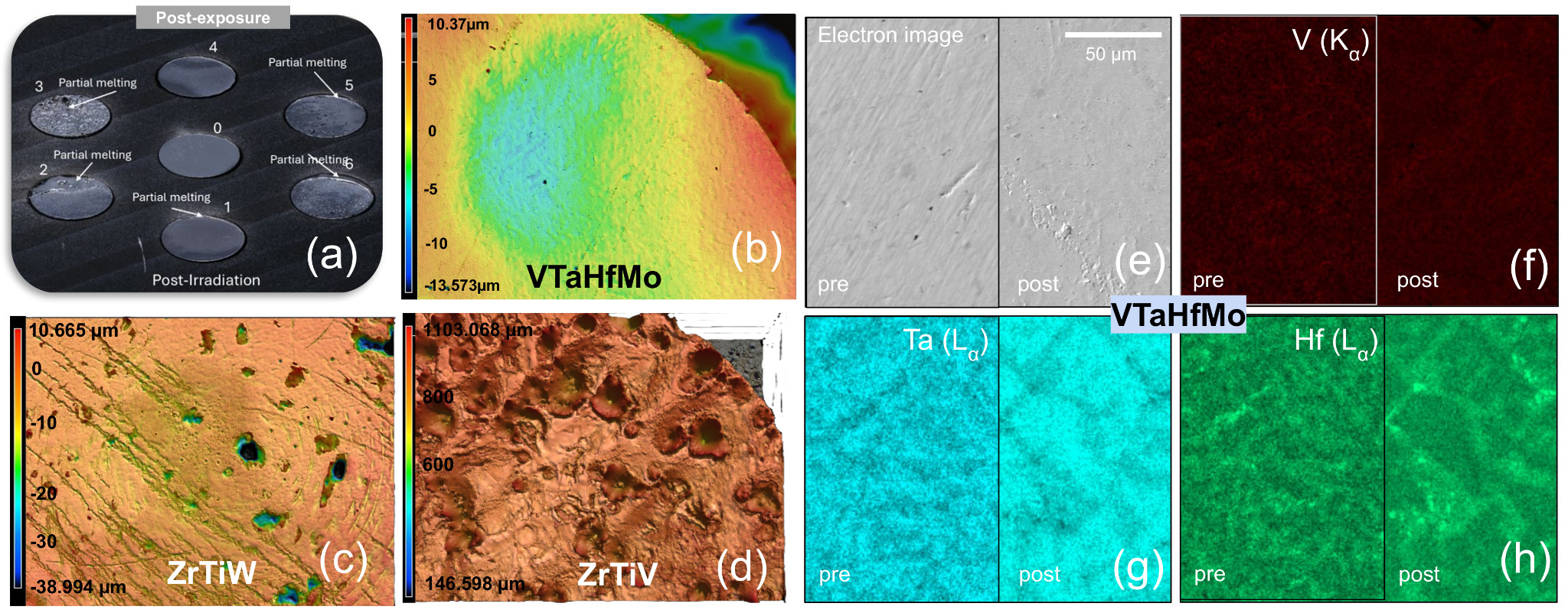}
\caption{Post-exposure characterization of refractory MPEAs following DIII-D plasma exposure. (a) Optical image of post-exposure alloy coupons showing localized partial melting after plasma exposure. (b--d) Representative 3D optical profilometry maps of exposed VTaHfMo, ZrTiW, and ZrTiV alloys, respectively; note the different height scales. (e) Secondary electron SEM images of the VTaHfMo alloy surface before and after plasma exposure. (f-h) Corresponding EDS elemental maps of the VTaHfMo alloy pre- and post-plasma exposure.}
\label{fig:11}
\end{figure*}

Seven refractory alloy buttons were exposed on the Avalanche DiMES head (\#9): VTaHfMo, ZrTiTaHf, ZrTiCr, ZrTiV, ZrTiHf, ZrTiW, and ZrTiTa, corresponding to sample positions 0--6 in Figure~\ref{fig:11} (a). These systems are promising since they span what would be a fully HCP solid solution and all the way to BCC stabilizing systems. In all of the ternary alloys, the BCC element addition was < 10 at\%. The ZrTiTaHf, ZrTiHf, ZrTiV, ZrTiCr, ZrTiW, and ZrTiTa alloys were synthesized using spark plasma sintering (SPS), while VTaHfMo alloy was synthesized via arc melting. Most of the SPS-fabricated Zr--Ti-based alloys showed significant porosity after the synthesis process. Therefore, the observed performance differences reflect both alloy chemistry and fabrication/surface-quality effects, rather than composition alone. The Avalanche DiMES head was exposed in seven rastered reference H-mode discharges, \#203772--203778, using flush button geometry, reaching peak heat fluxes of order \(\sim\)2--2.5 MW m\(^{-2}\). VTaHfMo showed the strongest overall response in this exposure set, with no major melting, no obvious surface damage, and only \(\sim\)10--40 \(\mu\)m-scale height variation by 3D profilometry. ZrTiW showed localized dendritic melt-like tracks, whereas ZrTiV showed the largest surface relief in the set (Figure~\ref{fig:11}(d)). EDS/LIBS mapping of VTaHfMo indicated no significant elemental loss or redistribution after exposure, although Hf segregation or refinement was observed and requires grain-boundary-scale analysis (Figure~\ref{fig:11} f-h). Several Zr/Ti-containing alloys (e.g. ZrTiCr, ZrTiV) showed stronger degradation, including voids, grooves, localized melting, compositional elemental segregation, and surface composition changes, with surface finish and edge heating likely contributing to the observed damage. In ZrTiW, dendritic surface tracks were observed, consistent with localized melt-like features in the 3D profilometry analysis (Figure~\ref{fig:11}(c)). Within this exposure set, VTaHfMo demonstrated the highest stability, whereas the stronger degradation observed in several Zr--Ti-based alloys was coupled to porosity, surface finish, local melting, and composition-dependent segregation.

\subsection{Chromium}\label{sec:cr}

To address core-edge integration challenges, chromium (Cr) was investigated as an alternative plasma-facing material to W, particularly for main-chamber PFCs where particle fluxes are lower than in the divertor. Cr exhibits several attractive properties as a plasma-facing material. It combines high thermal conductivity \cite{Holzwarth2002} (\(>90~\mathrm{W\,m^{-1}\,K^{-1}}\) at room temperature) with a relatively high melting point \cite{Arblaster2025} (2135~K), making it suitable for high-temperature applications. Furthermore, Cr can act as an effective scrape-off layer (SOL) radiator, similarly to Argon or Krypton, while producing core cooling rates more than an order of magnitude lower than those associated with W. Cr is also an efficient oxygen getter \cite{Dylla1986}, exhibits low neutron activation \cite{Zucchetti1996}, and demonstrates low hydrogen-isotope retention \cite{Yu2026}. Despite these potential advantages, the primary drawback of Cr is its lower erosion resistance compared with W \cite{Sugiyama2016}. Therefore, experimental characterization of Cr sputtering is essential for assessing the lifetime of Cr layers on the first wall of an FPP. However, data on Cr erosion in a tokamak environment is lacking. The present experiment therefore provides an initial tokamak measurement of Cr gross erosion under DIII-D divertor plasma conditions.

As a preliminary step in this direction, two Cr samples flush-mounted on the DiMES probe were exposed in DIII-D during the material thrust campaign [shots \#204933-\#204940], as described in Section~\ref{sec:micro-w}. The Cr samples experienced the same plasma conditions discussed in Section~\ref{sec:micro-w}, namely H-mode, strike-point rastering, and a peak inter-ELM heat flux of \(2~\mathrm{MW\,m^{-2}}\).

Inter-ELM Cr erosion rates were measured in-situ using the HRUV spectrometer \cite{Losada2021,Ennis2023}, whose line of sight was aligned with the DiMES center. In this configuration, only the central Cr sample was viewed by the HRUV spectrometer. Post-exposure Auger measurements indicated substantial cross-contamination between samples, with carbon, tungsten and oxygen being the dominant contaminants.  Photon fluxes from the Cr I multiplet at 357.87, 359.35, and 360.53 nm were converted to gross erosion rates using S/XB coefficients derived from ADAS data \cite{Summers2004}, with LS-coupled data split into J-resolved levels using ColRadPy. Here, the effective erosion yield denotes the spectroscopically inferred gross material-release flux normalized to the incident ion flux; it does not isolate physical sputtering from other release mechanisms. The inferred erosion rates were consistent across the three Cr I lines, with less than \(\sim\)5\% spread, and reached peak inter-ELM values of order 3--4 \(\times\) 10\(^{20}\) m\(^{-2}\) s\(^{-1}\) during strike-point sweeps in representative shot \#204937. Here, the spectroscopically inferred effective erosion yield represents the total atom release flux into the plasma (from physical sputtering and any thermal or surface-state contributions) normalized to the local incident ion flux.

\begin{figure}[htbp]
\centering
\includegraphics[alt={Measured chromium erosion yield versus electron temperature compared with RustBCA simulation predictions.},width=0.7\linewidth]{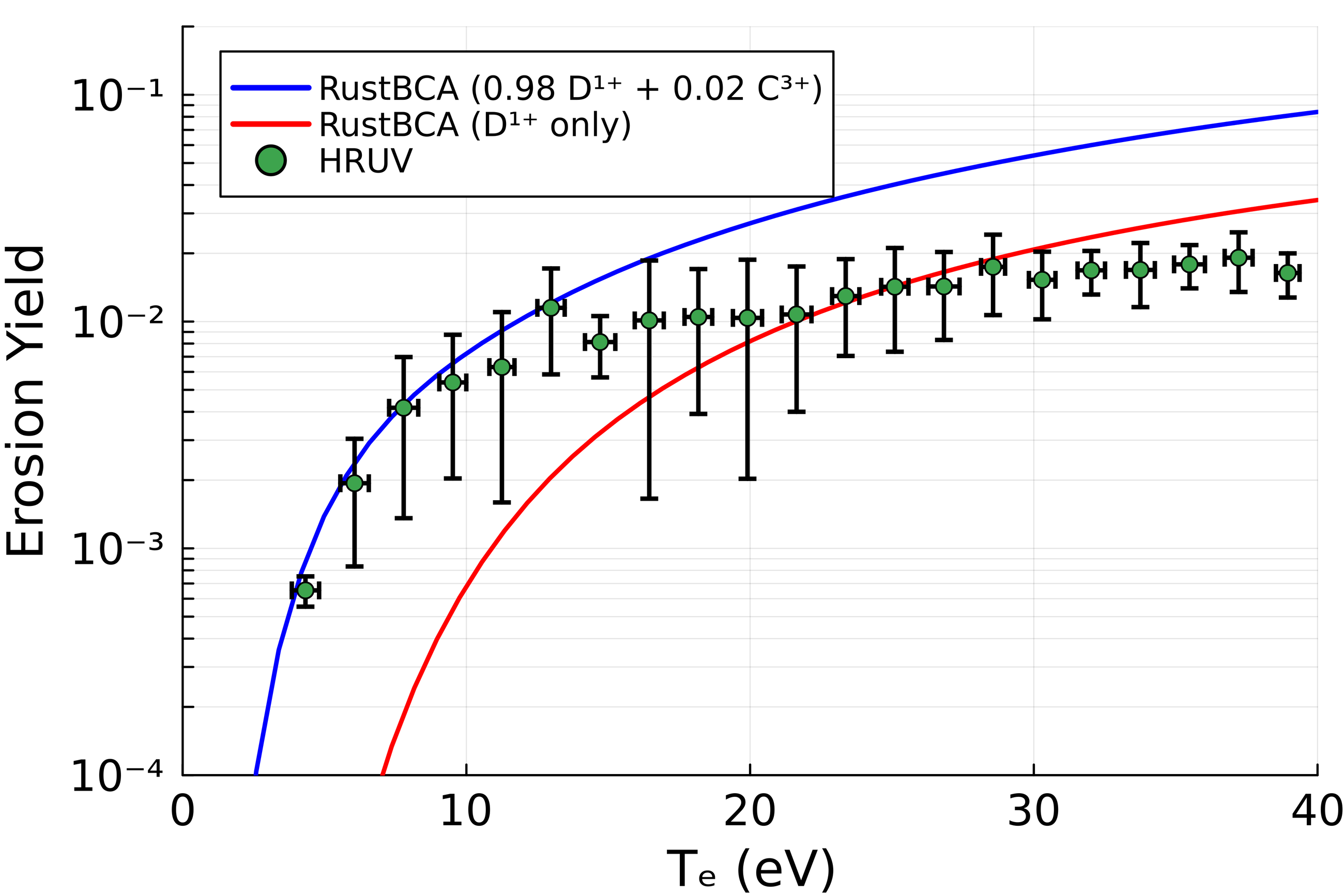}
\caption{Effective Cr erosion yield as a function of electron temperature, compared with RustBCA predictions \protect\cite{Drobny2021} using two plasma mixtures. The erosion yield was determined by dividing the erosion rate inferred with the HRUV spectrometer \protect\cite{Losada2021,Ennis2023} by the ion flux measured with Langmuir probes. The Cr I triplet at 357.87, 359.35, and 360.53~nm was used to infer the erosion rates.}
\label{fig:12}
\end{figure}

The measured Cr gross erosion rate followed the strike-point rastering waveform and remained approximately two orders of magnitude below the incident particle flux measured by Langmuir probes at the sample location (Figure~\ref{fig:01}). The inferred Cr erosion yield saturated at \(\sim (1\text{--}2)\times10^{-2}\) for target electron temperatures above \(\sim 10~\mathrm{eV}\) (Figure~\ref{fig:12}). Depending on the electron temperature at the target, the measured Cr erosion yield was found to be 2 to 3 times lower than predictions from RustBCA \cite{Drobny2021}, a binary-collision-approximation model for ion transport in matter, using as input a plasma mixture representative of erosion conditions at the lower outer strike point (OSP) of DIII-D \cite{Cappelli2026}. Further investigations are ongoing to understand this discrepancy. One possible explanation is near-surface Cr dilution caused by other contaminants.

This preliminary exposure of Cr to DIII-D divertor plasmas successfully achieved the objective of measuring the Cr erosion yield in a tokamak environment, which, to the authors' knowledge, had not previously been reported. The results provide an initial benchmark for Cr sputtering models and demonstrate the feasibility of dedicated PMI studies of Cr in DIII-D. While further work is needed to assess Cr layer lifetime and mixed-material evolution under main-chamber-relevant conditions, this study supports further evaluation of Cr as a candidate PFM for future fusion devices.

\section{Refractory ceramic and renewable plasma-facing concepts}

\subsection{Low-Z ceramic first-wall candidates}\label{sec:lowz}

Candidate dielectric ceramic plasma-facing materials provided by Helion Energy were exposed in DIII-D to evaluate erosion, morphology evolution, and survivability under diagnosed transient ELMy H-mode heat loading relevant to pulsed magneto-inertial fusion first-wall applications. Exposures were carried out using the standard H-mode OSP-rastered plasma configuration developed for the thrust campaign. Three low-Z ceramics were investigated: chemical-vapor-deposited silicon carbide (CVD SiC), silicon nitride (Si\(_{3}\)N\(_{4}\)), and boron carbide (B\(_{4}\)C).

Silicon nitride and silicon carbide have previously been proposed as candidate plasma-facing materials. Silicon nitride is characterized by high flexural and compressive strength, excellent thermal stability, and resistance to thermal shock \cite{Krstic2012}. Silicon carbide has been studied as a plasma-facing component for its capacity to maintain material integrity under high heat fluxes, for being a low radiator (low-Z material), and for showing smaller chemical erosion yields compared to pure graphite \cite{Rudakov2020,Abrams2021,Zamperini2023}. Boron carbide is of particular interest due to its high neutron absorption capability associated with the large neutron capture cross section, combined with the potential for high-temperature operation enabled by its elevated melting point \cite{Han2023}.

\begin{figure}[htbp]
\centering
\includegraphics[alt={Photographs of ceramic sample sets before and after plasma exposure with a radial and toroidal orientation indicated.},width=0.99\linewidth]{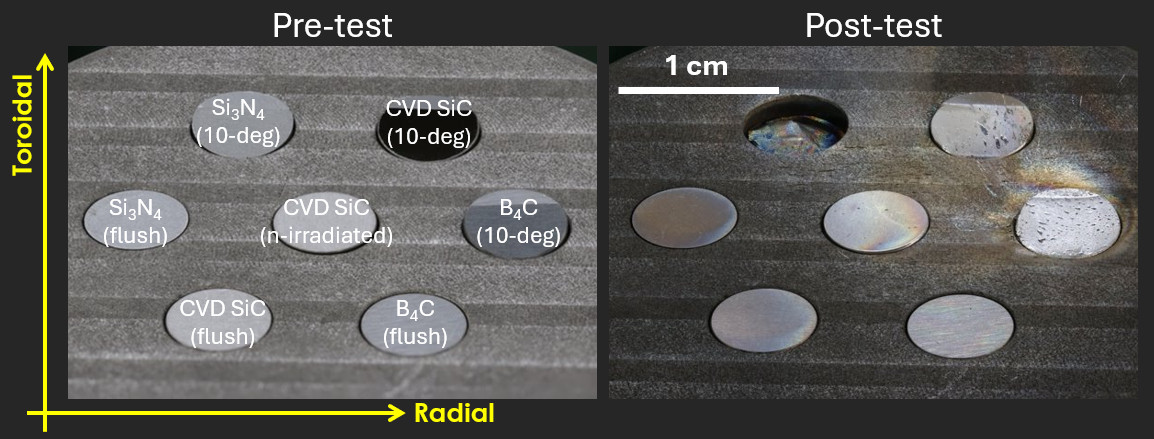}
\caption{Pre- and post-exposure photographs of DiMES-mounted ceramic first-wall candidate samples, including flush and 10\(^{\circ}\) angled B\(_{4}\)C, Si\(_{3}\)N\(_{4}\), and CVD SiC specimens, and one neutron-pre-irradiated CVD SiC specimen.}
\label{fig:13}
\end{figure}

Seven buttons were mounted in the DiMES \#5; four were flush-mounted, and three were raised and angled at 10 degrees towards the incident plasma fluxes to increase the impinging heat and particle flux; see Figure~\ref{fig:13}. One CVD SiC sample was neutron pre-irradiated before exposure to 6 \(\times\) 10\(^{-4}\) dpa.

\begin{figure}[htbp]
\centering
\includegraphics[alt={Effective inter-ELM silicon erosion yield versus target electron temperature inferred from Si II emission.},width=0.7\linewidth]{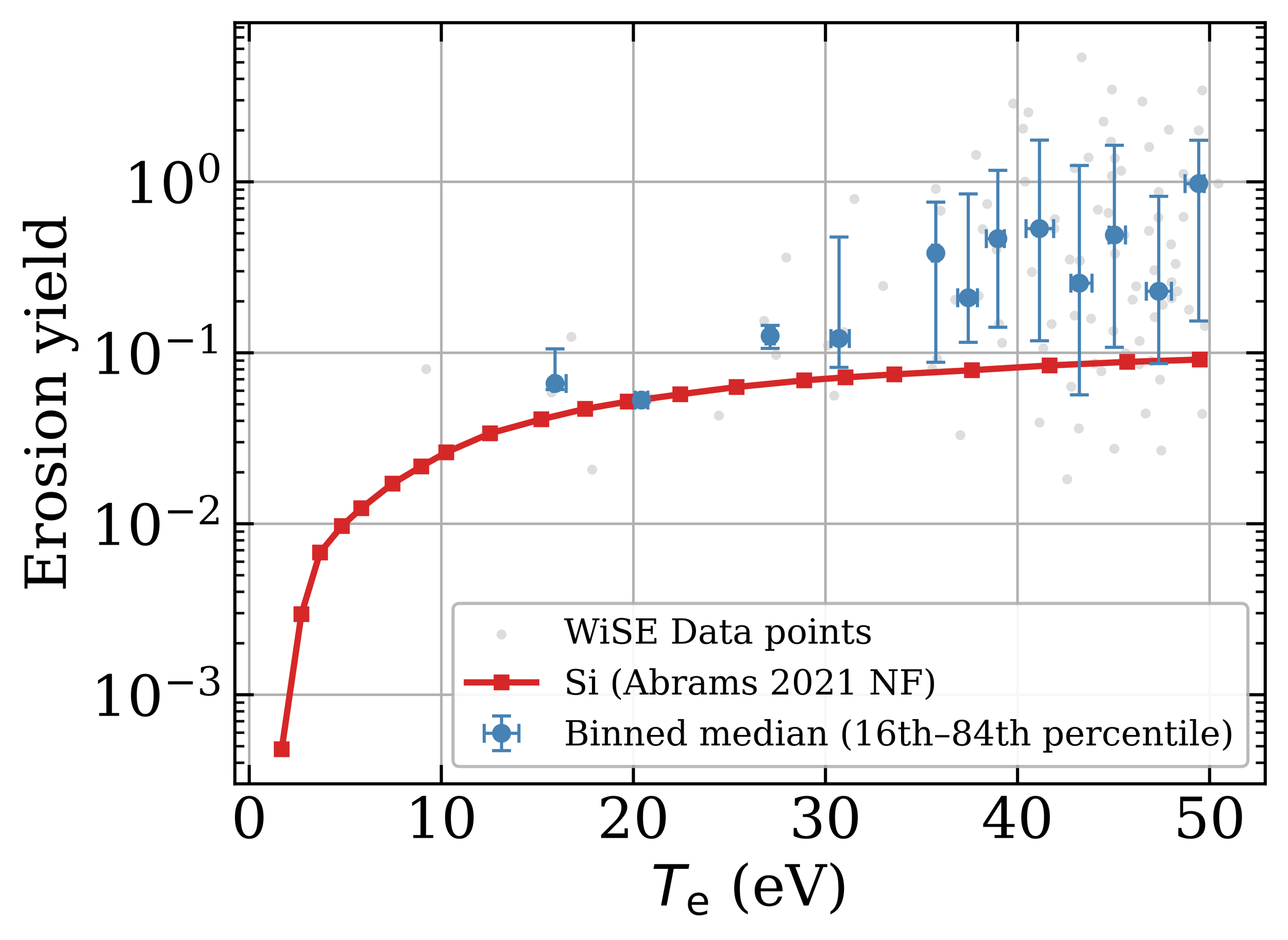}
\caption{Effective inter-ELM Si erosion yield inferred from WiSE measurements of Si II emission near 6347.1 \AA{} and 6371.4 \AA{} as a function of target electron temperature from Langmuir probe measurements. Error bars indicate the standard deviation within each temperature bin; intra-ELM emission was not resolved. Pure Si data from \cite{Abrams2021}.}
\label{fig:14}
\end{figure}

The samples were exposed in six successful H-mode discharges (\#203721- \#203726) with type-I ELMs at \(\sim\)65 Hz and strike-point rastering over \(\Delta R\) \(\approx\) 5 cm. The 10\(^{\circ}\) angled samples experienced transient ELM heat fluxes up to \(\sim\)40 MW m\(^{-2}\), with peak divertor plasma conditions of \(T_e\) \(\approx\) 50 eV and \(n_e\) \(\approx\) 5 \(\times\) 10\(^{18}\) m\(^{-3}\). The ELM decay time was \(\sim\)2 ms and \(\Delta W_{\mathrm{ELM}}/W_{\mathrm{MHD}}\) \(\approx\) 0.05. The inter-ELM ion flux at DiMES was estimated to be \(\sim\)10\(^{22}\) m\(^{-2}\) s\(^{-1}\), with peak intra-ELM heat and particle fluxes roughly an order of magnitude higher but lasting only \(\sim\)0.3 ms. The inter-ELM silicon erosion rate was measured in-situ using the Wide Spectral Emission (WiSE) multispectral region spectroscopy set \cite{McLean2019}, through analysis of the Si II emission near 6347.1 \AA{} and 6371.4 \AA{} using the unfocused WiSE view. Only inter-ELM Si emission was evaluated because intra-ELM photon signals could not be resolved in the available WiSE data, but may be significant under the \(\sim\)40 MW m\(^{-2}\) transient loading. As shown in Figure~\ref{fig:14}, the inferred effective Si erosion yield reached \(\sim\)0.5, dominated by emission from the angled SiC sample, and was approximately up to an order of magnitude larger than previous sputtering-yield measurements \cite{Abrams2021,Effenberg2023}. The origin of this discrepancy remains under investigation. One possible explanation is the contribution of additional Si evaporative losses associated with the expected high surface temperature reached by the tilted sample surfaces.

In addition to in-situ erosion measurements, a series of post-exposure characterizations were performed. B\(_{4}\)C exhibited partial surface melting and debris formation after exposure. Post-exposure compositional analysis indicated preferential B depletion, with the B/C ratio decreasing from 2.69 before exposure to 0.75, a \(\sim\)72\% reduction, after plasma exposure, leaving a C-enriched surface layer.

The angled silicon nitride specimen developed severe cracking and failed after the first discharge exposure (\#203721), with flying debris observed from the DiMES region at \(\sim\)2.9 s and \(\sim\)3.7 s, suggesting insufficient resistance to the imposed transient thermomechanical loading; thermal-expansion mismatch or thermal-shock response may have contributed. Under the present 10\(^{\circ}\) angled H-mode exposure conditions, Si\(_{3}\)N\(_{4}\) did not retain structural integrity and therefore appears unsuitable for this transient high-heat-flux first-wall application.

Among the tested low-Z ceramics, CVD SiC showed the best overall response, with the flush, angled, and neutron-pre-irradiated SiC specimens remaining macroscopically intact through the six H-mode discharges. However, in-situ spectroscopy and post-exposure analysis indicated preferential erosion of Si relative to C. Based on the current understanding, this behavior cannot be explained solely by sputtering processes, suggesting that additional mechanisms, such as thermally driven evaporation of Si, may have contributed to silicon release during plasma exposure.

Ultimately, CVD SiC emerged as the most resilient low-Z ceramic tested, but its high effective Si source and preferential Si loss motivate further studies of its viability as a first-wall material for pulsed fusion systems.

\subsection{Ultra-high temperature ceramics}\label{sec:uhtc}

TiB\(_{2}\) was evaluated in the HFIR-compatible DiMES exposure set under ohmic plasma conditions, while additional carbide-based UHTCs were tested separately as flush-mounted buttons under rastered H-mode exposure. The TiB$_2$-containing head was exposed during 16 Ohmic-heated discharges (\#203789--203804) in the $5^\circ$-inclined holder, with estimated sample-normal heat fluxes of $\sim0.8$--1.35~MW\,m$^{-2}$ (Section~\ref{sec:irr-w}).

Post-exposure SEM showed roughening on TiB\(_{2}\) but little evidence of blistering, whereas blistering was more apparent on the more outboard W-based samples, especially HFIR-W and K-doped W. No obvious cracking was identified in the examined TiB$_2$ SEM regions. TiB\(_{2}\) also exhibited the highest D\(_{2}\) desorption peak temperature, 849 \(^{\circ}\)C, compared with 710 \(^{\circ}\)C for ITER-W and 704 \(^{\circ}\)C for K-doped W, suggesting higher-temperature D release states in the ceramic. Additionally, the TiB\(_{2}\) showed the lowest total D\(_{2}\) release among the four specimens in the ohmic exposure set (1.28 \(\times\) 10\(^{-7}\) mol), compared with 2.05 \(\times\) 10\(^{-7}\) mol for ITER-grade W and 2.88 \(\times\) 10\(^{-7}\) mol for K-doped W (Figure~\ref{fig:10}).

Linear plasma device studies on $\mathrm{TiB_2}$ utilizing low energy \(\sim\)40 eV deuterium ions at varied target temperatures are consistent with these results, demonstrating the material's resistance to significant steady-state plasma surface modification under conditions where W-based materials have been shown to blister \cite{Nuckols2024,Baldwin2024}.

\begin{figure}[htbp]
\centering
\includegraphics[alt={SEM image of damage to the ZrC specimen and two optical camera images of ejected material emission.},width=\linewidth]{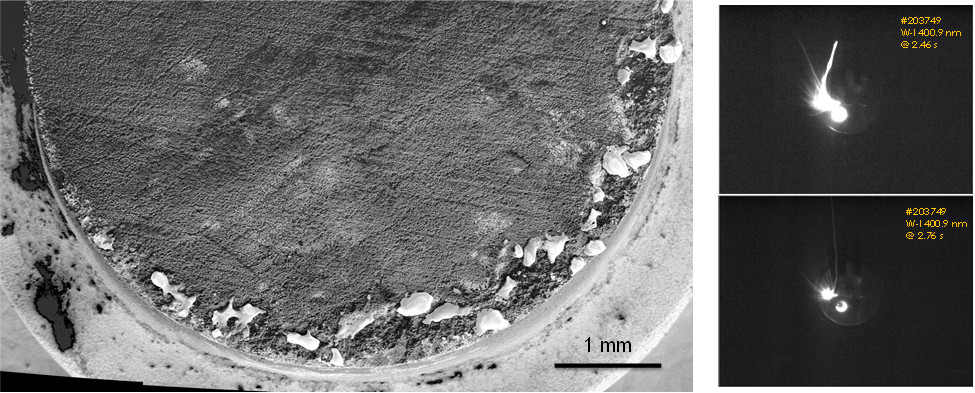}
\caption{Scanning electron microscope image of damage to a ZrC specimen created during plasma exposure. The captures on the right show emission from ejected material, observed by an optical camera with direct line-of-sight to the DiMES head.}
\label{fig:15}
\end{figure}

Additional carbide-based UHTCs were exposed under rastered H-mode conditions (\#203749--203755), using DiMES holders \#4 and \#14, with sample geometries listed in Table~\ref{tab:02}. The UHTC matrix in this campaign consisted of ZrC, NbC, and \((\mathrm{Nb}_{0.5}\mathrm{Ta}_{0.5})\mathrm{C}\). Additional contingency exposures included WC and $(\mathrm{W}_{0.5}\mathrm{Si}_{0.5})\mathrm{C}$. The measured densities were 5.30, 7.68, 10.59, and 15.60~g\,cm$^{-3}$ for ZrC, NbC, $(\mathrm{Nb}_{0.5}\mathrm{Ta}_{0.5})\mathrm{C}$, and WC, respectively. Quantitative mass-loss measurements showed that ZrC was the clear outlier, losing 7.049 mg, equivalent to $\sim 47.0\,\mu\mathrm{m}$ thickness loss (apparent net thickness-change rate $\sim 3.4\text{--}3.9\,\mu\mathrm{m\,s^{-1}}$ over $\sim 12\text{--}14\,\mathrm{s}$ nominal H-mode dwell). In comparison, NbC lost 0.027 mg ($\sim 124\,\mathrm{nm}$, $\sim 8.9\text{--}10.4\,\mathrm{nm\,s^{-1}}$), $(\mathrm{Nb}_{0.5}\mathrm{Ta}_{0.5})\mathrm{C}$ lost 0.019 mg ($\sim 63.4\,\mathrm{nm}$, $\sim 4.5\text{--}5.3\,\mathrm{nm\,s^{-1}}$), and an ITER-W reference exposed under similar conditions lost 0.010 mg ($\sim 18.3\,\mathrm{nm}$, $\sim 1.3\text{--}1.5\,\mathrm{nm\,s^{-1}}$). Across the campaign, most carbide specimens exhibited only subtle surface evolution by optical inspection, consistent with net erosion being small over the short DiMES exposure durations. However, one clear outlier was ZrC, which exhibited intense, localized material emission observed by DiMES TV beginning with the first discharge \#203749 (Figure~\ref{fig:15}).  Emission appeared concentrated at the sample periphery/leading edge and was accompanied by visibly roughened, damaged regions post-exposure, including large resolidified/melt-like features on the order of \(\sim\)200 \(\mu\)m. Subsequent discharges observed decreases in DiMES TV Zr signals, consistent with leading edge smoothing. SEM imaging shows only minor surface roughening in the sample center, far away from the leading edge effects.

Consistent with this interpretation, the Nb-based carbides (NbC and NbTaC) emerged with only minimal macroscopic change and limited microscopic evolution, with fiducial markers and near-surface morphology largely retained in pre-/post-exposure SEM comparisons. In contingency testing, WC and $(\mathrm{W}_{0.5}\mathrm{Si}_{0.5})\mathrm{C}$ similarly showed no obvious new failure mode attributable to tokamak plasma exposure at these conditions, supporting the broader conclusion that several UHTC candidates can tolerate DIII-D divertor H-mode exposure provided geometric edge effects are controlled.

In summary, the TiB\(_{2}\) exhibited low D\(_{2}\) release and limited blistering, while most carbide UHTCs survived flush H-mode exposure. Collectively, these results motivate follow-up UHTC exposures under more extreme conditions, potentially moving to angled sample exposure for enhanced heat loads more conducive to limiter PFCs.

\subsection{Renewable boron-based plasma-facing concepts}\label{sec:boron}

Renewable boron pebble rods were tested in DIII-D as the first in-tokamak divertor exposure of a boron-pebble renewable PFC concept, motivated by localized high-heat-flux exhaust regions where fixed PFCs face erosion, cracking, melting, and co-deposition limits. The pebble aggregate materials are an option to address the damage induced by the high heat flux on the divertor by reconstituting the surface of the PFC, which releases nearly intact pebbles upon exposure to plasma heat loads. The surface of PFCs would be replenished by extrusion of the pebble conglomerate, composed of pebbles previously exposed to the plasma and recovered from the floor of the vacuum vessel \cite{MartinezLoran2023}. The concept is relevant to Thea Energy's Eos stellarator divertor studies, where renewable low-Z boron surfaces are being explored for localized exhaust handling rather than full-vessel coverage. It can be applied to regions of high localized heat flux --like the outer strike point (OSP), and in regions of high slag deposition --like the inner strike point (ISP), to remove accumulated deposits. The pebble release rate can be tailored to meet specific heat or slag removal requirements by adjusting the strength of the inter-pebble binder \cite{MartinezLoran2024}. Pebbles can be chosen from almost any material, but this concept would enable the use of low-Z materials for better core performance. Previous works demonstrated intact pebble recovery during laser bench test experiments up to \(40~\mathrm{MW\,m^{-2}}\) on carbon (C) pebble rods \cite{MartinezLoran2023,MartinezLoran2024}, but boron pebbles are more suitable for this application due to the better core-plasma compatibility of B \cite{Jackson1991,Hino1997,Effenberg2022}, and lower tritium retention compared to C \cite{Annen1997,Abe2025,Martinez-Loran2025}.

\begin{figure}[htbp]
\centering
\includegraphics[alt={Boron pebble rod photos and plots of released boron transport and radiated signals during exposure.},width=\linewidth]{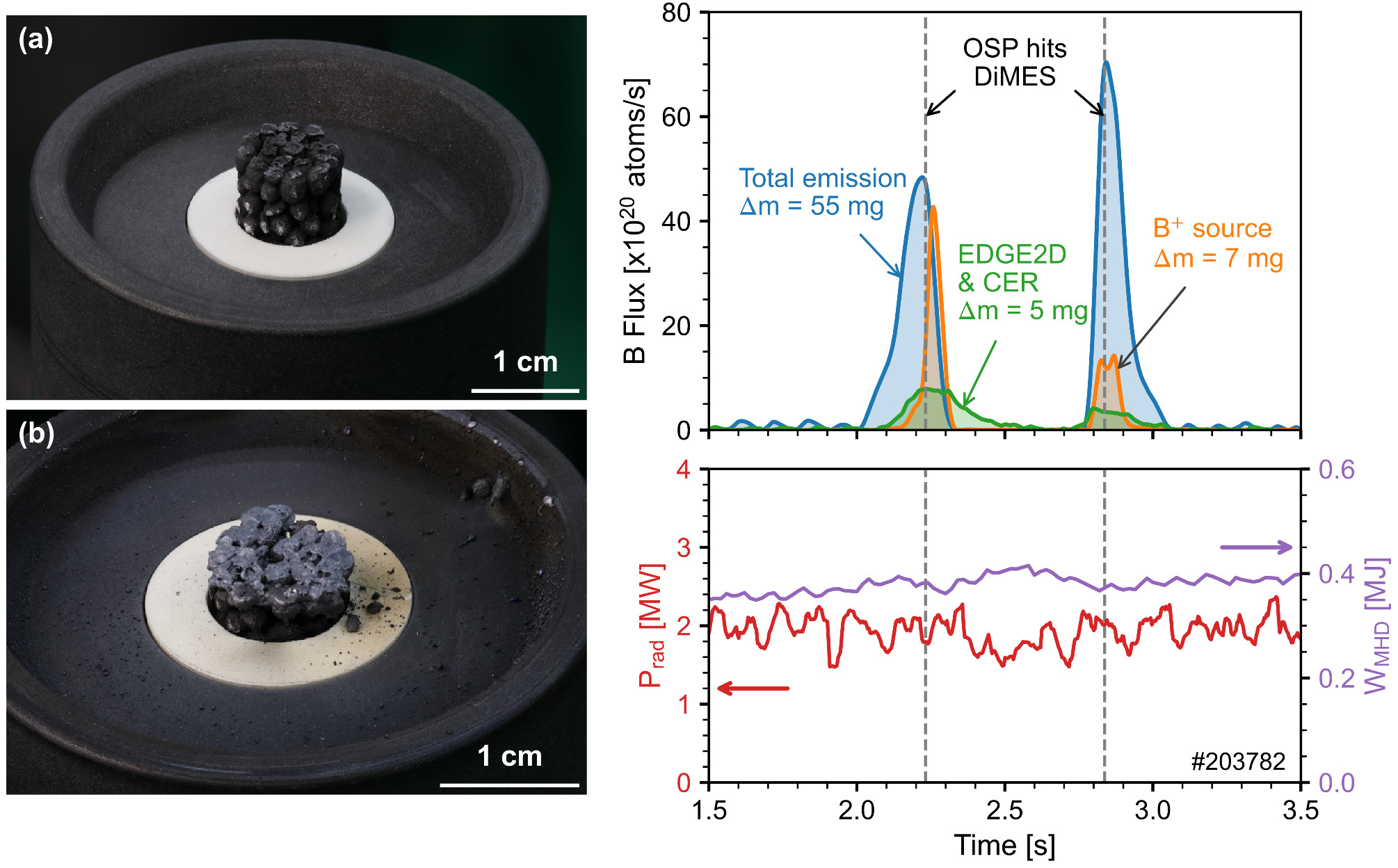}
\caption{Overview of boron pebble aggregate exposures to L-mode LSN divertor plasmas showing: (a) A sample of a boron pebble aggregate rod mounted on DiMES before the exposure. (b) The same sample after shot \#203781. (c) The boron emission rate from boron pebble rods during outer strike point sweeps in shot \#203782 with the total boron emission (blue), estimated ionization source in the divertor (orange), and the estimated divertor ionization source (green) from EDGE2D-EIRENE fits to core B\(^{5+}\) concentrations from CER measurements. (d) The radiated power (red), and core thermal energy (purple) during shot \#203782.}
\label{fig:16}
\end{figure}

Pebble aggregate rods of 1 cm diameter, produced from sintered amorphous B pebbles were exposed to incident heat loads up to \(q_{\parallel}=80~\mathrm{MW\,m^{-2}}\) in the DIII-D tokamak. The exposed rods included 2 mm and 5 mm protruding B pebble rods in shots \#203779--203785, including L-mode sweeps/dwells and a final downward VDE exposure. Single, fixed pebble rods were mounted on DiMES and exposed to L-mode lower single null (LSN) plasmas. Significant boron dust emission from the pebbles was observed and identified as the main source of ionized boron in the divertor. Figure~\ref{fig:16}(a) shows a boron pebble rod mounted on DiMES with an initial protrusion of 2 mm above shelf level. The same rod is shown in Figure~\ref{fig:16}(b) worn down after shot \#203781. Approximately half of the released aggregate was recovered locally in the DiMES cup as mm-scale fragments/pebbles, while the remaining material was not locally recovered. The source of ionized boron was attributed to dust ablation in the OSP. The fraction of ionized boron was estimated to be \(\sim\)13\% of the total boron emission, from the time-integrated boron emission rates during OSP sweeps shown in Figure~\ref{fig:16}(c). The total ionized boron emission shown in Figure~\ref{fig:16}(c) in orange, estimated from BII spectral imaging, agreed, within a factor of 1.5, with the time-integrated boron source ionization term estimated from EDGE2D fits to core B\(^{5+}\) levels from charge exchange recombination measurements (CER), shown in green. The remaining released boron was inferred to be lost predominantly as dust to the surrounding vessel. Closing the unrecovered boron balance requires 3D ablation/transport and erosion--redeposition modeling, as prior DIII-D localized boron material injection modeling showed nonlocal redistribution and mixed-material deposition \cite{Effenberg2025}.

Core performance was not significantly affected by the boron ionization source, as shown in Figure~\ref{fig:16}(d), where the radiated power (solid red) and the core thermal energy \(W_{\mathrm{MHD}}\) (solid purple) are observed to remain nearly constant during OSP sweeps. The experiment demonstrated tokamak compatibility of the renewable boron-pebble concept at the diagnostic and systems level, but amorphous sintered B pebbles produced excessive dust/particulate release and incomplete local recovery. As such, future efforts will be directed towards manufacturing pebbles from different forms of boron and optimizing the inter-pebble binder to the divertor heat loads, to tune the release rate for a reduced pebble duty cycle in the plasma and achieve nearly intact pebble recovery.

\section{Comparative assessment of plasma-facing material performance}
\label{sec:comparison}

\begin{figure*}[p]
\centering
\includegraphics[alt={Color-coded matrix summarizing cracking, erosion, melting and deuterium retention response for material classes in DIII-D.},
    width=0.92\textwidth,
    height=0.69\textheight,
    keepaspectratio
]{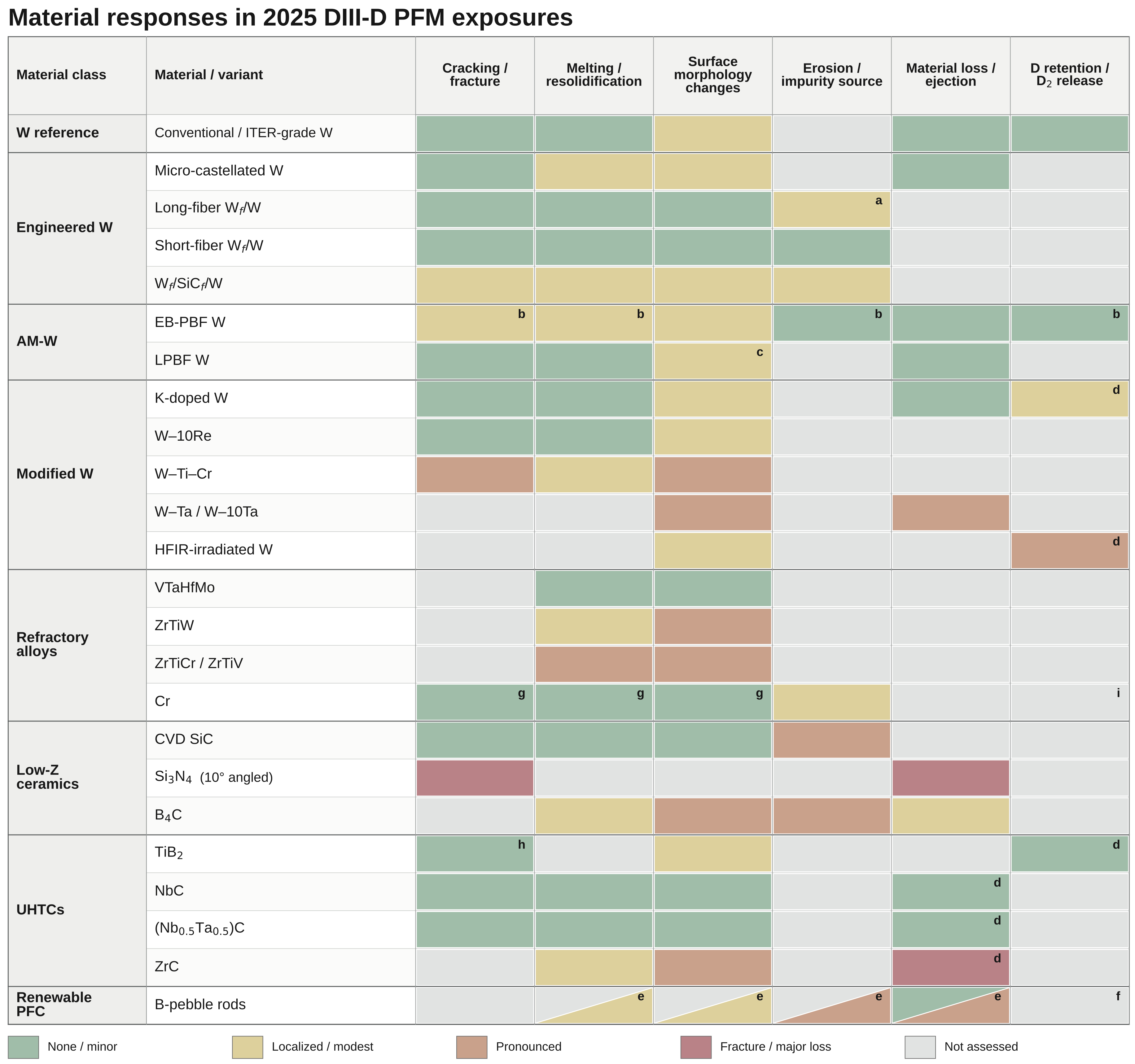}

\caption[Qualitative material responses in the 2025 DIII-D campaign]{Qualitative assessment of 2025 DIII-D PFM responses (Table~\ref{tab:03}) and selected complementary measurements. Green: none/minor; yellow: localized/modest; orange: pronounced; red: fracture/major loss; gray: not assessed. Colors indicate observed responses; exposure conditions and geometries differ. Lettered annotations indicate:

\textsuperscript{a} Long-fiber W$_{\mathrm f}$/W: fiber-arrested cracking.

\textsuperscript{b} EB-PBF W: DIII-D erosion and separate TPE retention measurements~\protect\cite{Zuniga2026EBPBF}.

\textsuperscript{c} LPBF W: additional morphology analysis to be reported elsewhere.

\textsuperscript{d} W/UHTCs: additional analyses will be reported elsewhere.

\textsuperscript{e} B-pebble rods: split cells distinguish the binder/inter-pebble matrix (upper-left) from B pebbles (lower-right). Gray denotes unresolved binder behavior; yellow localized pebble melting/morphology changes; orange B-dust emission
or substantial ejection. Green denotes intended aggregate shedding, with incomplete local recovery (up to $\sim$50\%)~\protect\cite{MartinezLoran2023,MartinezLoran2024}.

\textsuperscript{f} B-pebble D retention: separate BN-bound aggregates showed favorable fuel retention behavior in PISCES-A; not measured for the carbon-bound DIII-D
rods~\protect\cite{Martinez-Loran2025}.

\textsuperscript{g} Cr: no cracking/melting; minor arc tracks.

\textsuperscript{h} TiB$_2$: no obvious cracking in examined SEM regions.

\textsuperscript{i} Cr retention: separate PISCES-RF measurements \protect\cite{Yu2026}; DIII-D analysis to be reported elsewhere.
}
\label{fig:year2_material_response}
\end{figure*}

Table~\ref{tab:03} summarizes the material responses observed in the present DIII-D campaign. Figure~\ref{fig:year2_material_response} provides a color-coded qualitative assessment of the material responses summarized in Table~\ref{tab:03}, including selected complementary measurements. The classifications distinguish cracking, melting, surface modification, erosion, material loss, and D$_2$ retention/release. Tungsten remains the reference for high-heat-flux plasma-facing components because of its low sputtering yield, high melting point, comparatively high thermal conductivity, and lower fuel-retention propensity than carbon-based surfaces, despite its susceptibility to cracking, induced degradation, and the radiative impact of core W contamination \cite{Philipps2011,Brezinsek2015, Coenen2016DEMO,Linsmeier2017, Neu2026}. In the present tests, long-fiber W$_{\mathrm f}$/W exhibited crack arrest and micro-castellated W confined localized damage, while K-doped W, W--Re, and high-density AM-W largely retained macroscopic integrity. These findings are consistent with previous studies demonstrating that engineered W architectures can improve damage tolerance under transient thermal loading through fiber-mediated crack arrest and reduced near-surface thermal stresses \cite{Riesch2016,Terra2020}. In contrast, LPBF W--10Ta showed greater mass loss under angled exposure, and W--Ti--Cr developed extensive cracking under off-nominal loading. In contrast to the non-irradiated W screening in Year~1, neutron-irradiated W exhibited 2.8 times higher integrated D$_2$ release than pristine W under matched Ohmic exposures, with the main desorption peak shifting from $\sim710$ to $\sim630\,^{\circ}$C. Earlier DIII-D tests of dispersion-strengthened W likewise showed a strong dependence on composition and processing: W--TiC retained comparatively good matrix integrity, whereas W--TaC and W--ZrC exhibited dispersoid-related damage and melting under angled loading \cite{Coburn2026,Hargrove2026}.

Among the non-W refractory materials investigated here, VTaHfMo showed limited surface degradation relative to the other MPEAs. NbC and $(\mathrm{Nb}_{0.5}\mathrm{Ta}_{0.5})\mathrm{C}$ exhibited mass losses of 0.027 and 0.019~mg, respectively, compared with 0.010~mg for an ITER-W reference in related H-mode exposures. Cr may be relevant to lower-flux main-chamber surfaces, although its inferred erosion yield of $\sim(1\text{--}2)\times10^{-2}$ raises concerns about layer lifetime. Interpretation of some refractory-alloy results in the preceding DIII-D campaign was similarly limited by carbon-rich surface deposits \cite{Coburn2026}.

Lower-Z materials offer a smaller core-radiation penalty for impurity influx than W, but erosion, fuel retention, and thermomechanical integrity remain limiting considerations. Graphite is constrained by chemical erosion, tritium co-deposition, and neutron-induced property degradation \cite{Roth2009,Linke2019}. In the present campaign, Si$_3$N$_4$ fractured during the first discharge, while B$_4$C showed localized melting and a 72\% reduction in the measured B/C ratio. CVD SiC remained macroscopically intact, including a specimen pre-irradiated to only $6\times10^{-4}$~dpa, but its effective Si erosion yield reached $\sim0.5$, approximately 5--10 times earlier DIII-D Si sputtering-yield measurements; thermally driven Si release may have contributed. The preceding DIII-D campaign also reported intact CVD SiC and SiC$_{\mathrm f}$/SiC specimens, but arc tracks on the composite and leading-edge erosion of plasma-sprayed SiC \cite{Coburn2026}. Cracking and chipping of SiSiC limiter material in TEXTOR and flaking of B$_4$C coatings in Tore Supra provide additional examples of ceramic degradation \cite{Koizlik1986TEXTOR,Corre2011Flaking}. Furthermore, D-implantation studies found SiC retention comparable to or higher than graphite and B$_4$C saturation concentrations comparable to graphite, while Si-containing carbon codeposits also retained substantial D \cite{Mayer1998, Balden1999Codeposition, Koller2019,Lantaigne2022}.

TiB$_2$ showed the lowest integrated D$_2$ release in the matched Ohmic exposure set ($1.28\times10^{-7}$~mol versus $2.05\times10^{-7}$~mol for pristine W), although its higher desorption peak ($849\,^{\circ}$C) and limited heat-flux testing do not establish low reactor fuel retention. Renewable B-pebble rods provide a different approach to erosion management: approximately half of the released aggregate was recovered locally, but substantial particulate losses remained. Separate PISCES-A studies of BN-bound B-pebble aggregates demonstrated low D retention \cite{Martinez-Loran2025}, whereas D retention of the carbon-bound rods exposed here was not measured. The fate of unrecovered particulates remains an important concern. ITER solid B injection modeling predicts that millimeter-sized particles can penetrate beyond the LCFS before ablating under prescribed injection conditions \cite{Effenberg2026ITER}, which motivates the need to assess nonlocal B transport and redeposition, including associated fuel retention through B co-deposition \cite{Pitts2025}. The carbon binder requires further assessment in view of D trapping associated with B--C--D bonding in DIII-D boronization layers \cite{Abe2025}. Earlier DIII-D experiments demonstrated, with real-time material injection of B, BN, and Li, a complementary approach to supplying low-Z material for replenishable wall coatings and power exhaust \cite{Effenberg2022}. The Year~1 campaign demonstrated a complementary liquid-metal concept using Li-filled W capillary porous structures under H-mode loading, with preheating suppressing visible droplet ejection \cite{Coburn2026}. Sustained replenishment, material recovery, and fuel-inventory control remain to be established for these renewable PFC concepts.

These results support further evaluation of selected non-W materials for specific PFC applications, but do not establish direct replacements for W under reactor divertor loading. Quantitative comparisons remain most reliable within matched exposure sets; differences in sample geometry, processing, and local heat flux, together with short uncooled exposures and limited neutron-damage data, preclude a universal material ranking. Following DiMES coupon-scale screening, selected W-based materials, including EB-PBF and K-doped W, were advanced to divertor tile-scale testing during the 2026 DIII-D Tungsten Tile Testing (T3) campaign, alongside candidate W coating and cladding technologies relevant to ITER first-wall applications and the DIII-D full-wall changeout \cite{Coburn2026TOFE}. Post-exposure characterization is ongoing, with detailed results to be reported separately. Further qualification should prioritize engineered W architectures, additively manufactured W, and selected W alloys, using matched ITER-W references to quantify erosion, cracking, and D/T retention under repeated high-heat-flux loading and following neutron irradiation \cite{DOEFSTRoadmap2026,Pintsuk2026}. Progression toward component qualification requires actively cooled PFC mock-ups to assess manufacturing reproducibility, joints, and interfaces under cyclic thermal loading \cite{Pintsuk2026}. In parallel, renewable PFC concepts, including liquid-metal systems, warrant further development for localized high-heat-flux applications, particularly regarding sustained replenishment, material migration, and fuel-inventory control \cite{DOEFSTRoadmap2026,UKFusionMaterials2025}.

\clearpage
\begin{table*}[p]
\centering
\caption{Cross-material summary of reported DIII-D plasma-facing material responses. Comparisons across different exposure conditions are qualitative; mass loss, effective erosion yield, and integrated D$_2$ release denote distinct measurements.}
\label{tab:03}
\begingroup
\fontsize{7.4}{7.8}\selectfont
\setlength{\tabcolsep}{1.2pt}
\renewcommand{\arraystretch}{0.97}
\begin{tabular}{@{}
>{\raggedright\arraybackslash}p{1.45cm}
>{\raggedright\arraybackslash}p{2.55cm}
>{\raggedright\arraybackslash}p{1.95cm}
>{\raggedright\arraybackslash}p{4.35cm}
>{\raggedright\arraybackslash}p{4.25cm}
>{\raggedright\arraybackslash}p{2.4cm}@{}}
\toprule
\textbf{Class} & \textbf{Material} & \textbf{Scenario} & \textbf{Key result} & \textbf{Dominant response} & \textbf{Assessment} \\
\midrule

\textbf{W refs.}
 & ITER-grade W (Ohmic)
& \#13; $5^\circ$ angled
& D$_2$ release $2.05\times10^{-7}$ mol; peak $710\,^\circ$C
& Pristine W reference for D retention
& Ohmic reference \\
\addlinespace[0.25em]
 & ITER-grade W (W--Ta set)
& \#7--8; flush H-mode
& Mass loss of two separate flat W specimens: 0.0500 and 0.3350 mg
& Scatter between the flat ITER-W references
& Limits precision of W--10Ta/W ratio \\
\addlinespace[0.25em]
 & ITER-grade W (carbide set)
& Related H-mode set
& Mass loss 0.010 mg ($\sim$18.3 nm equivalent)
& Low mass loss in reference exposure
& W reference; geometry-dependent \\
\addlinespace[0.25em]
 & Conventionally manufactured W
& \#1--2; flush L/H-mode
& H-mode roughness increase $\sim$0.015~$\mu$m
& No major surface damage; C-associated EDS signal
& Comparable LPBF-W reference \\
\midrule
\textbf{Engineered W}
 & Micro-castellated W (two cut orientations)
& \#17; $10^\circ$ H-mode
& No macroscopic cracking or gross loss
& Localized edge-column melting; smaller features for one cut
& Promising; edge-sensitive \\
\addlinespace[0.25em]
 & Long-fiber W$_{\mathrm f}$/W
& \#17; $10^\circ$ H-mode
& No macroscopic failure
& W-matrix cracks arrested at fiber--matrix interfaces
& Best crack arrest among tested composites \\
\addlinespace[0.25em]
 & Short-fiber W$_{\mathrm f}$/W
& \#17; flush H-mode
& No visible cracking or melting
& Mild erosion; integrity retained
& Favorable at flush loading \\
\addlinespace[0.25em]
 & W$_{\mathrm f}$/SiC$_{\mathrm f}$/W
& \#17; flush H-mode
& Bulk region remained intact
& Localized leading-edge melting and resolidification
& Geometry-sensitive \\
\addlinespace[0.25em]
 & EB-PBF W, (001) texture
& \#15--16; flush/angled H-mode
& Grain ECD $71.6\pm51.8~\mu$m; density $>99.5\%$
& Roughening near pre-existing intergranular cracks
& Viable; defect-sensitive \\
\addlinespace[0.25em]
 & EB-PBF W, (111) texture
& \#15--16; flush/angled H-mode
& Grain ECD $73.8\pm41.7~\mu$m; density $>99.5\%$
& No macroscopic failure reported
& Viable; texture effects unresolved \\
\addlinespace[0.25em]
 & LPBF W ($\sim$97\% dense)
& \#1--2; flush L/H-mode
& H-mode roughness increase $\sim$0.015~$\mu$m
& Localized blister-like features; no macroscopic failure
& Viable; mild surface evolution \\
\midrule
\textbf{Modified W}
 & K-doped W
& \#3, \#7--8 H-mode; \#13 Ohmic
& D$_2$ release $2.88\times10^{-7}$ mol ($1.4\times$ pristine W); peak $704\,^\circ$C
& ITER-W-like morphology; increased D release
& Near ITER-W-like under H-mode \\
\addlinespace[0.25em]
 & Commercial W--10Re
& \#3; flush/angled H-mode
& No separate quantitative mass loss reported
& Grain-scale grooving; no major damage
& Near ITER-W-like \\
\addlinespace[0.25em]
 & W--Ti--Cr
& \#4; flush/angled H-mode
& Cracks $\sim$10--20~$\mu$m wide in angled sample
& Extensive cracking and localized melting; no fragmentation
& Severe damage; off-nominal $\sim14^\circ$ angle \\
\addlinespace[0.25em]
 & AM W--Ta (flush); LPBF W--10Ta (angled)
& \#7--8; H-mode
& Mass loss: 0.64 mg flat; 2.19--2.87 mg angled
& Roughening and raised coral-like features
& Greater loss for angled specimens \\
\addlinespace[0.25em]
 & Neutron-irradiated ITER-W (0.3 dpa, $550\,^\circ$C)
& \#13; $5^\circ$ Ohmic
& D$_2$ release $5.77\times10^{-7}$ mol ($2.8\times$ pristine W); peak $\sim630\,^\circ$C
& Increased D$_2$ release; altered desorption spectrum
& Consistent with irradiation-enhanced trapping \\
\midrule

\textbf{Refractory}
 & VTaHfMo MPEA
& \#9; flush H-mode
& Profilometry height variation $\sim$10--40~$\mu$m
& Limited surface change; no major melting
& Best response in tested MPEA set \\
\addlinespace[0.25em]
 & ZrTiW MPEA
& \#9; flush H-mode
& Localized dendritic, melt-like tracks
& Surface modification; melt-like features
& Strong degradation \\
\addlinespace[0.25em]
 & ZrTiCr and ZrTiV MPEAs
& \#9; flush H-mode
& Largest profilometric relief in the set
& Voids, grooves, localized damage and segregation
& Degraded; processing-sensitive \\
\addlinespace[0.25em]
 & HfZrTiTa, ZrTiHf and ZrTiTa MPEAs
& \#9; flush H-mode
& No separate quantitative values reported
& Variable response; SPS processing porosity
& Inconclusive; porosity-sensitive \\
\addlinespace[0.25em]
 & Cr
& \#17; flush H-mode
& Effective erosion yield $\sim(1$--$2)\times10^{-2}$ for $T_e\gtrsim10$ eV
& Gross Cr erosion $\sim(3$--$4)\times10^{20}$ m$^{-2}$s$^{-1}$
& Main-wall candidate; erosion-limited \\
\midrule
\textbf{Ceramics}
 & CVD SiC
& \#5--6, \#18; flush/angled H-mode
& Effective Si erosion yield $\sim$0.5 (angled); $\sim$5--10$\times$ earlier DIII-D Si yields
& Macroscopically intact; preferential Si release
& High Si source despite good structural integrity \\
\addlinespace[0.25em]
 & Neutron-pre-irradiated CVD SiC
& \#5; flush H-mode
& $6\times10^{-4}$ dpa; no macroscopic failure after six shots
& Integrity retained at low pre-irradiation dose
& Preliminary; low neutron dose \\
\addlinespace[0.25em]
 & Si$_3$N$_4$
& \#5; angled H-mode
& Failure after first discharge
& Severe cracking and ejected debris
& Failed at tested loading \\
\addlinespace[0.25em]
 & B$_4$C
& \#5; flush/angled H-mode
& B/C ratio $2.69\rightarrow0.75$ ($\sim72\%$ decrease)
& Localized melting; preferential B loss and debris
& Strong B depletion \\
\addlinespace[0.25em]
 & TiB$_2$
& \#13; $5^\circ$ Ohmic
& D$_2$ release $1.28\times10^{-7}$ mol; peak $849\,^\circ$C
& Roughening; limited visible blistering
& Lowest D$_2$ release in Ohmic set \\
\addlinespace[0.25em]
 & NbC
& \#4; flush H-mode
& Mass loss 0.027 mg ($\sim$124 nm equivalent)
& Limited surface evolution
& Low loss; promising \\
\addlinespace[0.25em]
 & (Nb$_{0.5}$Ta$_{0.5}$)C
& \#4; flush H-mode
& Mass loss 0.019 mg ($\sim$63.4 nm equivalent)
& Limited surface evolution
& Low loss at flush loading \\
\addlinespace[0.25em]
 & ZrC
& \#4; flush H-mode
& Mass loss 7.049 mg ($\sim$47.0 $\mu$m equivalent)
& Intense localized emission and leading-edge damage
& Severe edge degradation \\
\addlinespace[0.25em]
 & WC and $(\mathrm{W}_{0.5}\mathrm{Si}_{0.5})\mathrm{C}$
& \#14; flush H-mode
& No quantitative loss reported
& No obvious new macroscopic failure mode
& Initial integrity; further analysis needed \\
\midrule
\textbf{Renewable}
 & B-pebble rods (2/5 mm protrusion)
& \#10--12; L/H/VDE
& $\sim$13\% of released B ionized; $\sim$50\% of released aggregate recovered in DiMES cup
& Rod recession; B pebbles/fragments and dust released
& Incomplete local recovery of released B material \\

\bottomrule
\end{tabular}
\endgroup
\end{table*}
\clearpage

\clearpage

\section{Conclusion}

44 advanced plasma-facing materials from 12 institutions, including four public--private fusion partnerships, were exposed and comparatively assessed in DIII-D during the second 2025 material testing campaign to support FPP wall and divertor material down-selection in an integrated tokamak environment. Comparative testing used two new reference scenarios: a revised 2 Hz, \(\sim\)5 cm OSP-rastered H-mode scenario to broaden DiMES heat-flux profiles, and a low-power Ohmic scenario to expose HFIR-irradiated samples. The campaign differentiated candidate materials by dominant response mode, including crack arrest, recrystallization/roughening, erosion, D retention, localized melting, and material particulate transport.

Engineered W architectures retained macroscopic integrity under the tested flush and angled ELMy H-mode exposures, with damage largely confined to local melt/re-solidification regions rather than propagating into catastrophic cracking or bulk melting. Long-fiber $\mathrm{W_f/W}$ showed the clearest crack-arrest behavior at fiber interfaces, while micro-castellated W localized damage within segmented columns. $\mathrm{W_f/SiC_f/W}$ preserved overall integrity across most of the surface despite leading-edge melting and erosion.

High-density AM-W retained near-reference structural integrity under L-mode and ELMy H-mode exposure. The \(\sim\)97\% dense LPBF material showed minimal surface evolution and no major crack formation, with possible deuterium blistering observed on AM-W specimens. EB-PBF AM-W retained its grain texture but exhibited localized roughening and microcracking near pre-existing defects on angled samples, indicating that AM-W performance is influenced by microstructure and processing route.

W-Re and K-doped W retained near ITER-W-like behavior under high-heat-flux H-mode exposure, with no major macroscopic damage and comparatively low mass loss. In contrast, AM W-Ta showed stronger heat-flux-sensitive degradation, with apparent mass loss increasing from 0.64 mg for the flat sample to 2.19--2.87 mg for 10\(^{\circ}\) angled samples, together with roughening and raised coral-like surface features.

Neutron-irradiated ITER-grade W was exposed in DIII-D for the first time after 0.3 dpa irradiation in HFIR at 550 \(^{\circ}\)C, using low-power Ohmic exposures intended to limit defect annealing and benchmarked against pristine W, K-doped W, and TiB\(_{2}\). Neutron irradiation increased D retention by \(\sim\)2.8\(\times\) relative to pristine W, while post-exposure morphology was strongly influenced by local impurity deposition and particle-flow effects, producing spatially varying roughening, deposition, and blistering. TiB\(_{2}\) showed the lowest D\(_{2}\) release in the Ohmic exposure set.

Refractory MPEAs showed strongly composition-dependent stability under repeated H-mode exposure. VTaHfMo retained the best overall surface integrity, with minimal morphology evolution and no major melting, while Zr--Ti-based alloys showed more severe voiding, grooving, segregation, and localized melting.

Carbide UHTCs also showed strongly composition-dependent H-mode performance. ZrC was the clear outlier, with severe leading-edge emission/damage and \(\sim\)7 mg mass loss, while NbC and \((\mathrm{Nb}_{0.5}\mathrm{Ta}_{0.5})\mathrm{C}\) retained macroscopic integrity with only \(\sim\)0.02--0.03 mg mass loss. Nb-containing carbides were therefore the most stable UHTCs in this exposure set, while ZrC performance was limited by localized edge damage and likely geometry/porosity sensitivity.

Among ceramic first-wall candidates, CVD SiC retained macroscopic surface integrity with minimal visible damage, while Si\(_{3}\)N\(_{4}\) failed and B\(_{4}\)C showed localized melting and redeposition debris. Despite the strongest structural performance, SiC exhibited elevated effective Si erosion yields of \(\sim\)0.5, about 5--10\(\times\) above prior Si sputtering-yield measurements in DIII-D.

The first in-tokamak divertor exposure of a renewable boron pebble-rod PFC concept demonstrated plasma-driven recession under high heat flux, while the low ionized fraction of released boron (\(\sim\)13\%) and partial local recovery (up to \(\sim\)50\%) identified particulate transport and recovery control as the main integration challenges for further development of the concept.

Initial in-situ tokamak gross-erosion measurements yielded effective erosion yields of order 10\(^{-2}\), providing a first benchmark for Cr erosion modeling and motivating assessment of mixed-material evolution and Cr layer lifetime under main-chamber-relevant plasma conditions.

Within the tested conditions, engineered W architectures, high-density AM-W, W-Re/K-doped W, VTaHfMo, Nb-containing carbides, and CVD SiC showed the strongest class-specific performance, while AM W-Ta, Zr--Ti-based MPEAs, ZrC, Si\(_{3}\)N\(_{4}\), B\(_{4}\)C, and boron pebble rods exhibited clearer degradation or integration limitations. These integrated PMI results provide comparative benchmarks for PFM down-selection across tokamak, stellarator, compact, and pulsed fusion concepts, supporting further qualification through tile-scale testing, including the 2026 DIII-D tungsten tile campaign, and future AI/ML-assisted PFM development.

\section{Acknowledgements}

This material is based upon work supported by the U.S. Department of Energy, Office of Science, Office of Fusion Energy Sciences, using the DIII-D National Fusion Facility, a DOE Office of Science user facility, under Awards DE-AC02-09CH11466, DE-SCL0000109, DE-SCL0000110, DE-AC05-00OR22725, DE-FC02-04ER54698, DE-FG02-07ER54917, DE-SC0023378, DE-AC52-07NA27344. Sandia National Laboratories is a multimission laboratory managed and operated by National Technology \& Engineering Solutions of Sandia, LLC, a wholly owned subsidiary of Honeywell International Inc., for the U.S. Department of Energy's National Nuclear Security Administration under contract DE-NA0003525. The authors gratefully acknowledge the use of the equipment provided by the FEM and CF at the University of Birmingham.

The work was performed as a part of the U.S.-Japan PHENIX Cooperation Project on Technological Assessment of Plasma Facing Components for DEMO Reactors, supported by the U.S. Department of Energy (DOE), Office of Science, Fusion Energy Sciences and Ministry of Education, Culture, Sports, Science and Technology, Japan. This research used resources at the HFIR, a DOE Office of Science User Facility operated by ORNL.

\section{Disclaimer}

This report was prepared as an account of work sponsored by an agency of the United States Government. Neither the United States Government nor any agency thereof, nor any of their employees, makes any warranty, express or implied, or assumes any legal liability or responsibility for the accuracy, completeness, or usefulness of any information, apparatus, product, or process disclosed, or represents that its use would not infringe privately owned rights. Reference herein to any specific commercial product, process, or service by trade name, trademark, manufacturer, or otherwise does not necessarily constitute or imply its endorsement, recommendation, or favoring by the United States Government or any agency thereof. The views and opinions of authors expressed herein do not necessarily state or reflect those of the United States Government or any agency thereof.

\section*{Data availability}
The data that support the findings of this study are available from the corresponding author upon reasonable request.

\bibliographystyle{elsarticle-num}
\bibliography{references}
\end{document}